\documentclass[a4paper,fleqn]{cas-sc}

\usepackage{moreverb}
\usepackage{mathtools}
\usepackage{physics}
\usepackage{tabu}
\usepackage{multirow}
\usepackage{multicol}
\usepackage{bm}

\usepackage[numbers,sort&compress]{natbib}

\def\tsc#1{\csdef{#1}{\textsc{\lowercase{#1}}\xspace}}
\tsc{WGM}
\tsc{QE}
\tsc{EP}
\tsc{PMS}
\tsc{BEC}
\tsc{DE}

\begin{document}
\let\WriteBookmarks\relax
\def\floatpagepagefraction{1}
\def\textpagefraction{.001}

\shorttitle{Insensitivity-induced potential non-uniqueness of Bouc-Wen parameters}

\shortauthors{Kundu and Mukhopadhyay}

\title [mode = title]{Insensitivity-induced potential non-uniqueness in system identification of Bouc-Wen models} 

\author[1]{Adrita Kundu}

\ead{adkundu@iitk.ac.in}

\credit{Conceptualization, Methodology, Formal analysis, Investigation, Software, Validation, Visualization, Writing - Original draft preparation}

\affiliation[1]{organization={Indian Institute of Technology Kanpur},
    addressline={Kalyanpur}, 
    city={Kanpur},
    postcode={UP-208016}, 
    country={India}}

\author[1]{Suparno Mukhopadhyay}[orcid=0000-0003-2693-762X]

\ead{suparno@iitk.ac.in}  

\cormark[1]

\credit{Conceptualization, Methodology, Resources, Supervision, Writing - review \& editing}

\cortext[cor1]{Corresponding author}

\begin{abstract}
Under earthquakes, structural systems often exhibit nonlinear hysteretic behaviour and undergo damage. For structural health monitoring in such cases, accurate estimation of dynamic responses, structural parameters, hysteretic energy dissipation, etc., is required. However, in practical scenarios, during system identification of structural systems with the popular Bouc-Wen (BW) hysteretic restoring force mechanisms, the BW parameters may often be incorrectly estimated. For the same model, the estimated BW parameters may be different for different sets of input-output measurements, indicating potential non-uniqueness in the parameter estimates. Nonetheless, the non-unique and incorrectly estimated BW parameters may still result in dynamic responses and hysteretic restoring force-deformation behaviours, which are very similar to those obtained for the correct system. The estimated damage is also very similar with the incorrectly estimated BW parameters. In this work, the existence of alternate sets of BW parameters, which result in hysteretic restoring force behaviour similar to the true system, is studied analytically. Approximate expressions for the rate of change of the hysteretic force with deformation are derived and analyzed in detail. It is shown that alternate sets of BW parameters with significant deviations from a set of "true" BW parameters may exist, which result in the rate of change of the hysteretic force, and consequently, the restoring force behaviour and the dynamic responses, to remain very similar to that obtained with the "true" BW parameters. The existence of these alternate parameters results in potential non-uniqueness of the BW parameter estimates, despite satisfying analytical identifiability requirements. It is also shown that these alternate systems exhibit very similar inelastic response ($C_R$) spectra as well as damage estimates (quantified herein through the Park-Ang damage index). When the magnitudes of the "true" BW parameters are either too high or too low, the deviations of the alternate BW parameters can be even larger. Furthermore, the ranges of the alternate BW parameters also depend on the extent of the hysteretic action being developed by the input excitation. The results are illustrated using different inputs: sinusoidal, El Centro motion, and a suite of ground motions compatible with the Kanai-Tajimi spectrum. The results of this work help in a better understanding of the potential non-uniqueness issues associated with the estimation of the BW parameters from measured responses using any system identification technique, which is caused by the insensitivity of these parameters towards the dynamic responses of the structure.
\end{abstract}


\begin{highlights}
\item Highlighting issues with vibration-based system identification of Bouc-Wen parameters.
\item Novel analysis of the difference between hysteretic behaviours of two systems.
\item Existence of alternate Bouc-Wen parameters producing similar hysteretic behaviours.
\item Dependence of alternate Bouc-Wen parameters on their magnitudes and input excitation.


\end{highlights}


\maketitle

\section{Introduction}

When structural systems are subjected to earthquakes, their force-deformation behaviours often display non-linear characteristics. For structural health monitoring and damage assessment purposes, accurate modelling and identification of the non-linearities involved in the system dynamics are required. The phenomenon of hysteresis is one of the most common non-linearity present in the force-deformation behaviours of structural systems. Being caused due to cyclic load reversals during earthquakes or other extreme events, when the structure or a part of it is pushed into the inelastic zone repeatedly, hysteretic force-deformation behaviour is often observed~\cite{baber1985random}. Structural systems displaying hysteresis behaviour include base isolation and other vibration control devices, and structures made of masonry, wood, steel and reinforced concrete.

The Bouc-Wen (BW) model is a widely adopted phenomenological model for capturing hysteretic behaviour, offering significant versatility in representing a vast variety of hysteresis phenomena in numerous applications~\cite{bouc1967forced,wen1976method}. Modified with the ability to incorporate degrading as well as pinching effects as well~\cite{baber1985random,baber1986modeling}, the BW hysteresis model can efficiently represent complex hysteretic natures of structural members and their subsequent identifications~\cite{sues1988systems,ma2004parameter,goda2009probabilistic}. 

In the literature, numerous works have successfully utilized BW-based models for vibration-based system identification and estimation of dynamic responses of experimental structures displaying hysteretic behaviours under seismic excitation. In Yang et al.~\cite{yang2014damage}, a two-story Reinforced Concrete (RC) frame is subjected to a sequence of earthquake loads, and system identification and damage assessment are performed while considering plastic hinges to be modelled with the BW model. Erazo and co-workers have illustrated system identification applications with a hysteretic seismic protection device modelled with BW restoring force mechanism~\cite{erazo2018bayesian} and a full-scale RC seven-story shear wall structure modelled with a coupled elastic cantilever beam and non-linear BW chain system~\cite{erazo2019bayesian}.  Chatzi et al.~\cite{chatzi2010experimental} have used the BW model along with other basis functions for identifying the hysteretic behaviour of a nonlinear non-conservative dissipative joint element. BW-based modelling and subsequent system identification have also been widely adopted for seismic base isolators~\cite{constantinou1990teflon,calabrese2018adaptive,niola2019nonlinear,ojha2022application,brewick2025hierarchical}, wire–cable vibration isolators~\cite{ni1998identification,xie2013identification,niu2021dynamic}, Magnetorheological dampers~\cite{spencer1997phenomenological,miah2015nonlinear,zhu2019efficient} and bolted joints~\cite{lin2022identification,teloli2021bayesian}. In Li and Wang~\cite{li2021parameter} and Kundu and Mukhopadhyay~\cite{kundu2025truncated}, system identifications are performed on a four-storied lab-scale steel frame and a four-storied lab-scale RC frame, respectively, with both being modelled with BW non-linear shear chain system. Cheng and Becker~\cite{cheng2021performance} carried out system identification of triple friction pendulum and lead rubber bearing having BW models. Apart from these, numerous other works have also utilized BW-based models to represent the hysteretic force-deformation of structural systems subjected to extreme events, such as earthquakes, as well as to perform subsequent system identification and damage assessment.

An important aspect for performing system identification of non-linear structures is the observability/identifiability of the numerical model vis-à-vis the available response measurements. It has to be ensured that, given no measurement noise and no modelling errors, only a unique set of parameters can produce the responses measured with an applied input excitation. Analytical treatment of the identifiability for the BW model has been previously studied~\cite{ikhouane2007systems,chatzis2015observability}. Sufficient richness of the input excitation is also required for ensuring identifiability. However, in practical scenarios, due to the presence of measurement noise, modelling errors and other factors, the requirements for analytical identifiability are necessary but not sufficient to ensure uniqueness in the estimated structural parameters with any system identification method. Therefore, in practical scenarios with a given set of dynamic measurements from the structure, despite the analytical identifiability requirements being fulfilled for a BW-based structural model, the parameter estimates of the model may not always be accurate and may instead exhibit non-uniqueness. In this work, the term "potential non-uniqueness" is used to address this phenomenon. Hernandez and Erazo~\cite{hernandez2024lower} have shown with numerical examples that, in the cases when the extent of hysteresis action is very low and the damage incurred is very minor, the estimates of the BW parameters may not be reliable. In Kundu and Mukhopadhyay~\cite{kundu2025truncated}, joint state-parameter estimations were carried out for numerical four degrees of freedom (DOFs) hysteretic chain models, from noisy acceleration measurements, utilizing Unscented Kalman filter (UKF)-based algorithms. It was observed that, while the estimates of dynamic states (displacements and velocities), hysteretic restoring force behaviours, and the stiffness and damping parameters were highly accurate, the estimates of the parameters governing the BW hysteretic behaviours deviated significantly from their true values. Potential non-uniqueness was observed in the estimates of the BW parameters, with different estimates obtained for different sets of noisy input-output measurements, for the same BW model.

With this potential non-uniqueness indicating insensitivities of the BW parameters towards dynamic responses, a sensitivity analysis was performed in Kundu and Mukhopadhyay~\cite{kundu2025truncated} for structural systems having hysteretic BW elements, with each BW element being governed by three shape parameters only. The analysis was performed by varying each structural parameter one at a time and quantifying the resulting changes in the dynamic responses, where it was observed that the three BW parameters were, in fact, sensitive towards the dynamic responses. These observations are also corroborated by other one-parameter-at-a-time sensitivity studies~\cite{ma2004parameter,dadheech2024estimation}. Therefore, requiring further investigation for the insensitivities of the BW parameters, it was finally observed in Kundu and Mukhopadhyay~\cite{kundu2025truncated} that the incorrectly estimated alternate sets of BW parameters actually resulted in very similar rates of change of the BW hysteretic force with deformation, which led to very similar hysteretic force-deformation behaviour and dynamic responses being produced by the alternate BW parameters. This work focuses on understanding the aforementioned potential non-uniqueness in the BW parameter estimates, which arise primarily due to the insensitivity of the BW parameters towards the dynamic responses of the structure.

In this work, the existence of alternate BW parameters, which produce very similar rates of change of the BW hysteretic restoring force with deformation, is studied. Considering the BW hysteresis model without any degradation and pinching effects, the alternate BW parameters are created by varying a "true" set of BW parameters by a range of values. The resulting differences between the rates of change of hysteretic force of the alternate and "true" BW parameters are analyzed. Regions of alternate parameters are found, which vary from the "true" parameters by substantial magnitudes and yet result in very similar rates of change of hysteretic force. The results are illustrated numerically using sinusoidal and El Centro ground motion inputs. It is also verified that such alternate BW parameters result in similar restoring force behaviours, dynamic responses, and damage estimates quantified through the Park-And damage index. Further, the alternate parameters also result in very similar inelastic displacement ratio spectra. When system identifications of such systems are performed with any identification technique, the alternate BW parameters may be captured instead of the original BW parameters, resulting in the potential non-uniqueness of the BW parameter estimates. This is demonstrated in this work through joint state-parameter estimations of a numerical four DOFs hysteretic chain system, which is subjected to a suite of ground motions compatible with a Kanai-Tajimi type spectrum.


\section{Single DOF system with Bouc-Wen restoring force element}\label{Section: Single DOF}

A single DOF (SDOF) lumped mass system having a BW restoring force element, and subjected to external force $f(t)$, is considered. The governing equation of motion of the system is given as:
\begin{equation}{\label{SDOF EOM}}
    m\Ddot{y} + c\dot{y} + \alpha ky + (1-\alpha)D_ykr = f(t)
\end{equation}
where $y$, $\dot{y}$ and $\ddot{y}$ are the displacement, velocity and acceleration of the lumped mass, respectively; $m,k,c$ and $\alpha$ are the mass, stiffness, viscous damping coefficient and the ratio of post-yield stiffness to initial stiffness, respectively; and $r$ is the hysteretic deformation of the BW element, whose evolution is governed by the following differential equation:
\begin{equation}{\label{BW evolution pre}}
    \dot{r} = \frac{1}{D_y}\left(A\dot{y} - \beta|\dot{y}|{|r|}^{n-1}r - \gamma\dot{y}{|r|}^{n} \right)
\end{equation}
where $D_y$ is the yield displacement of the BW element, and $A$, $\beta$, $\gamma$ and $n$ are the parameters governing its hysteretic force-deformation behaviour. In this work, only the thermodynamically admissible and realistic BW model ($-\beta\leq\gamma\leq\beta$, $n>1$) is considered, where $r\in\left[-r_{\textrm{max}},r_{\textrm{max}}\right]$ and $r_{\textrm{max}} = \left(\beta + \gamma\right)^{-\frac{1}{n}}$~\cite{ikhouane2007dynamic}. However, there is a redundancy in the parameter set $\{\alpha,k,A,\beta,\gamma,n\}$ for uniquely determining the system response and hysteretic behaviour~\cite{ma2004parameter}. Therefore, for ensuring the analytic identifiability of the BW parameters, the parameter $A$ is kept constant at $A=1$, as done in previous works~\cite{ma2004parameter,chatzis2015observability}. This makes the evolution of $r$ to be described as:
\begin{equation}{\label{BW evolution}}
    \dot{r} = \frac{1}{D_y}\left(\dot{y} - \beta|\dot{y}|{|r|}^{n-1}r - \gamma\dot{y}{|r|}^{n} \right)
\end{equation}
The hysteretic energy dissipated in the BW element due to $f(t)$ is given as:
\begin{equation}{\label{hyst_ener}}
    E_h = \int_{0}^{t}D_ykr\dot{y}\,dt \quad ,\text{for}\ j=1,\ldots,4
\end{equation}

For ease of analysis, $\dot{r}$ is separated into two different branches pertaining to different regions:
\begin{equation}{\label{BW branches}}
    \dot{r}=
\begin{cases}
    \frac{1}{D_y}\left(\dot{y} - \left(\beta + \gamma\right) \dot{y}|r|^{n}\right) \quad &,\text{for}\ \left(\dot{y}>0,\ r>0\right)\ \&\ \left(\dot{y}<0,\ r<0\right) \qquad \textrm{(I)} \\
    \frac{1}{D_y}\left(\dot{y} + \left(\beta - \gamma\right) \dot{y}|r|^{n}\right) \quad &,\text{for}\ \left(\dot{y}<0,\ r>0\right)\ \&\ \left(\dot{y}>0,\ r<0\right) \qquad \textrm{(II)}
\end{cases}
\end{equation}
Eq.~\ref{BW branches} is further modified to obtain: 
\begin{equation}{\label{dr_dy}}
    r^\prime_D = D_y\dv{r}{y}=
\begin{cases}
    1 - \left(\beta + \gamma\right)\left|r\right|^{n} \quad &,\text{for}\ \left(\dot{y}>0,\ r>0\right)\ \&\ \left(\dot{y}<0,\ r<0\right) \qquad \textrm{(I)} \\
    1 + \left(\beta - \gamma\right)\left|r\right|^{n} \quad &,\text{for}\ \left(\dot{y}<0,\ r>0\right)\ \&\ \left(\dot{y}>0,\ r<0\right) \qquad \textrm{(II)}
\end{cases}
\end{equation}
As the hysteretic restoring force developed in the BW element is $f_r=\left(1-\alpha\right)D_ykr$, the term $r^\prime_D = D_y\dv*{r}{y}$ of Eq.~\ref{dr_dy} is simply a scaled rate of change of the hysteretic part of the restoring force with the displacement of the lumped mass.

\section{Potential non-uniqueness of Bouc-Wen parameters: Alternate parameters giving similar hysteretic behaviour}\label{Section: Alternate BW}

Corresponding to the "true" BW parameters $\beta$, $\gamma$ and $n$, let us define an alternate set of BW parameters: $\bar{\beta}$, $\bar{\gamma}$ and $\bar{n}$. Let the two sets be related as: $\bar{\beta} = \beta + \Delta\beta, \bar{\gamma} = \gamma + \Delta\gamma$ and $\bar{n} = n + \Delta n$. In this section, we will show that $r^\prime_D$, and therefore, the hysteretic force-deformation behaviour as well, can be quite similar for the parameter sets $\{\beta,\gamma,n\}$ and $\{\bar{\beta},\bar{\gamma},\bar{n}\}$, even for substantial magnitudes of the differences $\Delta\beta$, $\Delta\gamma$ and $\Delta n$. 

First, we consider branch (I) of $r^\prime_D$ in Eq.~\ref{dr_dy}. For the parameter sets $\left\{\beta,\gamma,n\right\}$ and $\left\{\bar{\beta},\bar{\gamma},\bar{n}\right\}$, $r^\prime_D$ for branch (I) is denoted as $r^\prime_D(\textrm{I})$ and $\bar{r}^\prime_D(\textrm{I})$, respectively, and the difference $\bar{r}^\prime_D(\textrm{I}) - r^\prime_D(\textrm{I})$ is denoted as $\varepsilon_1(r)$. Therefore, $\varepsilon_1(r) = \left(\beta + \gamma\right)\left|r\right|^{n} - \left(\bar{\beta} + \bar{\gamma}\right)\left|r\right|^{\bar{n}}$. Considering $f(r) = \left(\beta + \gamma\right)\left|r\right|^{n}$, we have $\varepsilon_1(r) = f(r) - \left(\bar{\beta} + \bar{\gamma}\right)\left|r\right|^{\bar{n}}$. Taking the logarithm of $f(r)$ and $f(r) - \varepsilon_1(r)$, we get:
\begin{align}
    \log{f(r)} &= n\log{|r|} + \log{\left(\beta+\gamma\right)} \label{log f} 
    \\ 
    \nonumber \\
    \log{\left(f(r) - \varepsilon_1(r)\right)} &= \bar{n}\log{|r|} + \log{\left(\bar{\beta}+\bar{\gamma}\right)} \nonumber \\
    \implies \log{f(r)} + \log{\left(1 - \frac{\varepsilon_1(r)}{f(r)}\right)} &= \left(n + \Delta n\right)\log{|r|} + \log{\left(\beta+\gamma\right)} + \log{\left(1 + \frac{\Delta\beta + \Delta\gamma}{\beta + \gamma}\right)} \label{log f minus eps}
\end{align}
For $\bar{r}^\prime_D(\textrm{I})$ to be close to $r^\prime_D(\textrm{I})$, $\varepsilon_1(r)$, and thus, $\frac{\varepsilon_1(r)}{f(r)}$ should be small quantities, making the term $\log{\left(1 - \frac{\varepsilon_1(r)}{f(r)}\right)}\approx - \frac{\varepsilon_1(r)}{f(r)}$. Using this approximation in Eq.~\ref{log f minus eps} and subtracting Eq.~\ref{log f} from it, an approximation of $\varepsilon_1(r)$ is obtained as:
\begin{equation}{\label{epsilon 1}}
    \varepsilon_1(r) \approx -\left(\beta + \gamma\right)\left|r\right|^{n}\left(\Delta n\log{|r|} + \log{\left(1 + \Delta_1\right)}\right)
\end{equation}
where $\Delta_1 = \frac{\Delta\beta + \Delta\gamma}{\beta + \gamma}$. The goodness of this approximation is illustrated later with a numerical example. It is also implied from Eq.~\ref{epsilon 1} that $(1+\Delta_1)>0$ always. The properties of $\varepsilon_1(r)$ in Eq.~\ref{epsilon 1} and the metrics used to quantify it are now studied.

\begin{figure}[h!]
  \centering
    \includegraphics[width=0.4\textwidth]{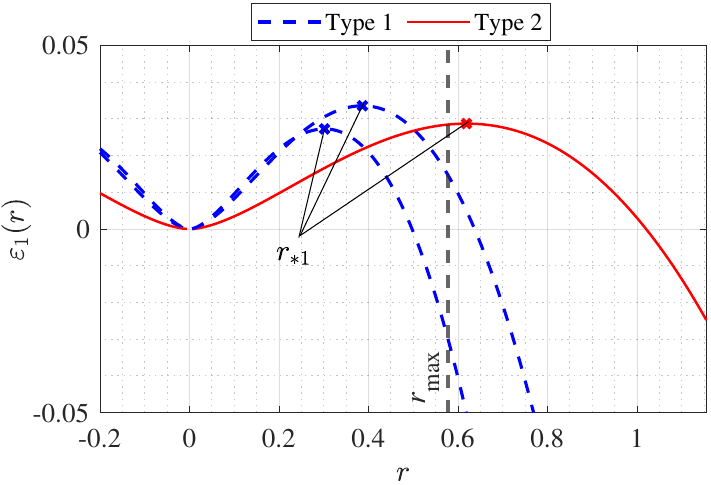}
  \caption{Typical curves of $\varepsilon_1(r)$ for $\Delta n>0$.}
  \label{Fig: Typical eps_1}
\end{figure}

As $\varepsilon_1(r)$ in Eq.~\ref{epsilon 1} is an even function in $r$, for ease of analysis, its properties are evaluated for $r\in\left[0,r_{\text{max}}\right]$. $\varepsilon_1(r)$ is smooth and continuous $\forall r\geq 0$, where its stationary points, obtained from $\dv{\varepsilon_1}{r}=0$, are at $r=0$ and at $r = r_{*1} = \exp{-\left(\frac{1}{n} + \frac{\log{\left(1+\Delta_1\right)}}{\Delta n}\right)}$. The values of $\varepsilon_1(r)$ at $r=0$ and $r_{*1}$, are $0$ and $\frac{\Delta n\left(\beta + \gamma\right)}{ne\left(1 + \Delta_1\right)^{\frac{n}{\Delta n}}}$, respectively. Also, $\dv[2]{\varepsilon_1}{r}\Bigr|_{r=r_{*1}} = -n\Delta n\left(\beta + \gamma\right)e^{-1 + \frac{2}{n}}\left(1 + \Delta_1\right)^{\frac{2-n}{\Delta n}}$. It is inferred that when $\Delta n>0$, $\varepsilon_1(r)$ monotonically increases in $0\leq r<r_{*1}$, has a positive maxima at $r=r_{*1}$, and then monotonically decreases for $r>r_{*1}$. Similarly, when $\Delta n<0$, $\varepsilon_1(r)$ monotonically decreases in $0\leq r<r_{*1}$, has a negative minima at $r=r_{*1}$, and then monotonically increases for $r>r_{*1}$. Typical curves of $\varepsilon_1(r)$ for $\Delta n>0$ are shown in Fig.~\ref{Fig: Typical eps_1}. Two types of curves are obtained, denoted as type 1 and type 2 curves. For type 1 curves, the extreme values of $\varepsilon_1(r)$ in $r\in\left[0,r_{\text{max}}\right]$, are at $r=r_{*1}$ and $r_{\text{max}}$, while for type 2 curves, the extreme values are only at $r=r_{\text{max}}$. For physical meaningfulness, these extreme values of $\varepsilon_1(r)$, i.e., of the deviation of $\bar{r}^\prime_D(\textrm{I})$ from $r^\prime_D(\textrm{I})$, are normalized with the average value of $r^\prime_D(\textrm{I})$, denoted by $r_{D,\textrm{avg}}^\prime$, as:
\begin{equation}{\label{eps1 metrics}}
    \epsilon_1 = \frac{\varepsilon_1(r_{\text{max}})}{r_{D,\textrm{avg}}^\prime} = \frac{1}{r_{D,\textrm{avg}}^\prime}\log\frac{\left(\beta + \gamma\right)^{\Delta_n}}{\left(1 + \Delta_1\right)}; 
    \quad \textrm{and} \quad
    \epsilon_{*1} = \frac{\varepsilon_1(r_{*1})}{r_{D,\textrm{avg}}^\prime} = \frac{\Delta_n\left(\beta + \gamma\right)}{r_{D,\textrm{avg}}^\prime e\left(1 + \Delta_1\right)^{\frac{1}{\Delta_n}}}
\end{equation}
where $\Delta_n = \frac{\Delta n}{n}$ and $r_{D,\textrm{avg}}^\prime$ is calculated by dividing the area of $r^\prime_D(\textrm{I})$, denoted by $A_{r^\prime_D(\textrm{I})}$, by $r_{\textrm{max}}$:
\begin{equation}{\label{rD_prime_area}}
    A_{r^\prime_D(\textrm{I})} = \int_0^{r_{\text{max}}}\left(1 - \left(\beta + \gamma\right)\left|r\right|^{n}\right)\dd r = \frac{n}{\left(\beta+\gamma\right)^{1/n}\left(n+1\right)}; \quad r_{D,\textrm{avg}}^\prime = \frac{A_{r^\prime_D(\textrm{I})}}{r_{\textrm{max}}} = \frac{n}{n+1}
\end{equation}
Furthermore, the total absolute deviation of $\bar{r}^\prime_D(\textrm{I})$ from $r^\prime_D(\textrm{I})$, denoted by $A_{\varepsilon_1}$, is quantified by computing the total area of $\varepsilon_1(r)$, and then normalizing with $A_{r^\prime_D(\textrm{I})}$ as:
\begin{equation}{\label{eps1 area}}
    A_{\varepsilon_1} = \frac{1}{A_{r^\prime_D(\textrm{I})}}\int_0^{r_{\text{max}}}\abs{\varepsilon_1(r)}\dd r
\end{equation}
The three metrics: $A_{\varepsilon_1}$, $\epsilon_{*1}$ and $\epsilon_1$ are used as a measure of the deviation of $\bar{r}^\prime_D(\textrm{I})$ from $r^\prime_D(\textrm{I})$.


The branch (II) of $r^\prime_D$ in Eq.~\ref{dr_dy} is also examined in a similar manner. With $r^\prime_D(\textrm{II})$ and $\bar{r}^\prime_D(\textrm{II})$ denoting $r^\prime_D$ for branch (II), for $\left\{\beta,\gamma,n\right\}$ and $\left\{\bar{\beta},\bar{\gamma},\bar{n}\right\}$, respectively, their difference is given as: $\varepsilon_2(r) = \bar{r}^\prime_D(\textrm{II}) - r^\prime_D(\textrm{II})$, and approximated as:
\begin{equation}{\label{epsilon 2}}
    \varepsilon_2(r) \approx \left(\beta - \gamma\right)\left|r\right|^{n}\left(\Delta n\log{|r|} + \log{\left(1 + \Delta_2\right)}\right)
\end{equation}
where $\Delta_2 = \frac{\Delta\beta - \Delta\gamma}{\beta - \gamma}$, and $(1+\Delta_2)>0$ always. $\varepsilon_2(r)$ is very similar to $\varepsilon_1(r)$ of Eq.~\ref{epsilon 1} in all aspects except the sign of the function. For the considered class of Bouc-Wen model i.e., $-\beta\leq\gamma\leq\beta$ and $n>1$, when $\Delta n<0$, $\varepsilon_1(r)<0$ and $\varepsilon_2(r)>0$ initially. Similarly, when $\Delta n>0$, $\varepsilon_1(r)>0$ and $\varepsilon_2(r)<0$ initially. Therefore, the typical curves of $\varepsilon_2(r)$, when $\Delta n<0$, can also be represented by the curves in Fig.~\ref{Fig: Typical eps_1}, where its stationary point for $r>0$ is $r_{*2} = \exp{-\left(\frac{1}{n} + \frac{\log{\left(1+\Delta_2\right)}}{\Delta n}\right)}$. Considering $\varepsilon_2(r)$ for $r\in\left[0,r_{\text{max}}\right]$, its extreme values are: $\varepsilon_2(r_{\text{max}})$ and/or $\varepsilon_2(r_{*2})$. Similar to the Eq.~\ref{eps1 metrics}, these extreme values are also normalized with $r_{D,\textrm{avg}}^\prime$ of Eq.~\ref{rD_prime_area} to obtain:
\begin{equation}{\label{eps2 metrics}}
    \epsilon_2 = \frac{\varepsilon_2(r_{\text{max}})}{r_{D,\textrm{avg}}^\prime} = \frac{1}{\kappa r_{D,\textrm{avg}}^\prime}\log\frac{\left(1 + \Delta_2\right)}{\left(\beta + \gamma\right)^{\Delta_n}}; 
    \quad \textrm{and} \quad
    \epsilon_{*2} = \frac{\varepsilon_2(r_{*2})}{r_{D,\textrm{avg}}^\prime} = \frac{-\Delta_n\left(\beta - \gamma\right)}{r_{D,\textrm{avg}}^\prime e\left(1 + \Delta_2\right)^{\frac{1}{\Delta_n}}}
\end{equation}
where $\kappa = \frac{\beta + \gamma}{\beta - \gamma}$. The net deviation of $\bar{r}^\prime_D(\textrm{II})$ from $r^\prime_D(\textrm{II})$ is denoted by $A_{\varepsilon_2}$, and is obtained similarly to $A_{\varepsilon_1}$ in Eq.~\ref{eps1 area}, as:
\begin{equation}{\label{eps2 area}}
    A_{\varepsilon_2} = \frac{1}{A_{r^\prime_D(\textrm{I})}}\int_0^{r_{\text{max}}}\abs{\varepsilon_2(r)}\dd r
\end{equation}
The three metrics: $A_{\varepsilon_2}$, $\epsilon_{*2}$ and $\epsilon_2$ are used as a measure of the deviation of $\bar{r}^\prime_D(\textrm{II})$ from $r^\prime_D(\textrm{II})$.

For a given set of $\left\{\beta,\gamma,n\right\}$, substantial $\Delta\beta$, $\Delta\gamma$ and $\Delta n$ can be found which result in low values of the metrics: $A_{\varepsilon_1}$, $\epsilon_{*1}$, $\epsilon_1$, $A_{\varepsilon_2}$, $\epsilon_{*2}$ and $\epsilon_2$, which indicate low deviations between the $r^\prime_D$ (in both branches (I) and (II)) for $\left\{\beta,\gamma,n\right\}$ and $\left\{\bar{\beta},\bar{\gamma},\bar{n}\right\}$. This also means high similarities between the hysteretic behaviours of the true and alternate parameter sets. This is shown in the numerical illustration next.

\subsection{Numerical illustration}

\begin{figure}[h!]
  \centering
    \includegraphics[width=0.8\textwidth]{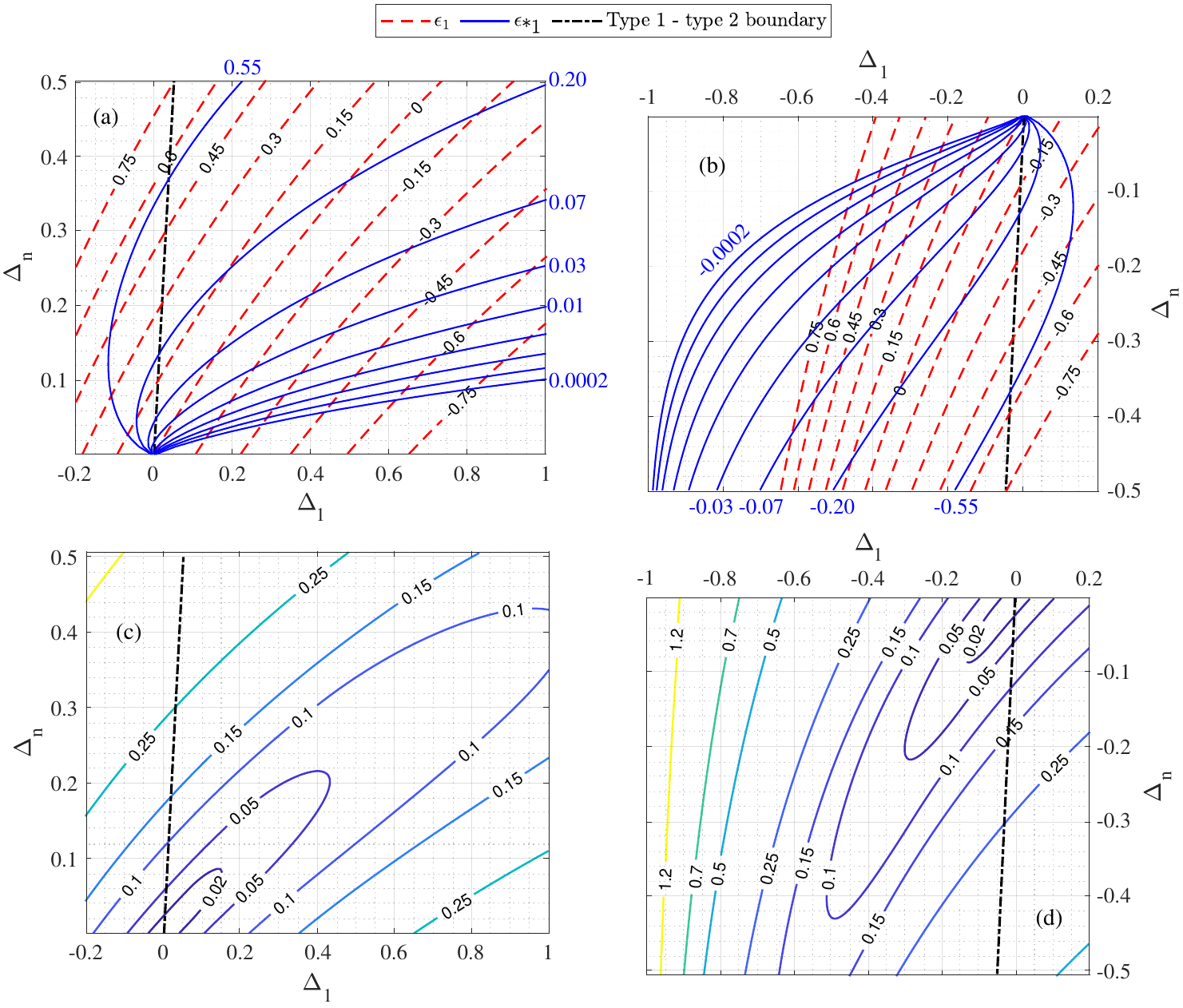}
  \caption{Contour plots of $\varepsilon_1(r)$ metrics for $\beta = 2, \gamma = 1$ and $n = 2$: (a-b) $\epsilon_1$ and $\epsilon_{*1}$, and (c-d) $A_{\varepsilon_1}$.}
  \label{Fig: varepsilon_1 contours}
\end{figure}

Considering $\beta = 2, \gamma = 1$ and $n = 2$, the deviations of $\bar{r}^\prime_D(\textrm{I})$ and $\bar{r}^\prime_D(\textrm{II})$ from $\bar{r}^\prime_D(\textrm{I})$ and $\bar{r}^\prime_D(\textrm{II})$, respectively, for numerous alternate BW parameters, are observed. First, the contour plots of the metrics $\epsilon_1$, $\epsilon_{*1}$ and $A_{\varepsilon_1}$, are obtained for $\Delta_n\in\left(-0.5,0.5\right]$ and $\Delta_1 \in\left(-1,1\right]$, and shown in Fig.~\ref{Fig: varepsilon_1 contours}. The boundary demarcating type 1 and type 2 curves, obtained by equating $r_{\text{max}}=r_{*1}$, is also shown. In Figs.~\ref{Fig: varepsilon_1 contours}a and c, where $\Delta_n>0$, the regions to the right of the boundary belong to type 1 curves, whereas, in Figs.~\ref{Fig: varepsilon_1 contours}b and d, where $\Delta_n<0$, the regions to the left of the boundary belong to type 1 curves. The values of $\epsilon_{*1}$ are only meaningful in the region of type 1 curves. The following observations are drawn from Fig.~\ref{Fig: varepsilon_1 contours}:
\begin{enumerate}
    \item $\epsilon_1$ and $A_{\varepsilon_1}$ are lower in the region of type 1 curves. 
    \item The signs of $\epsilon_{*1}$ are different for Figs.~\ref{Fig: varepsilon_1 contours}a and b, as $\varepsilon_1(r_{*1}) = \frac{\Delta_n\left(\beta + \gamma\right)}{e\left(1 + \Delta_1\right)^{\frac{1}{\Delta_n}}} >0$, when $\Delta n>0$, and $\varepsilon_1(r_{*1})<0$, when $\Delta n<0$.
    \item As $(1 + \Delta_1)>0$, all the contours extending to $\Delta_1\leq 0$ in Figs.~\ref{Fig: varepsilon_1 contours}b and d are squeezed in $\Delta_1\in \left(-1,0\right]$. This leads to relatively lower magnitudes of $\Delta_1$, for $\Delta_1<0$ than $\Delta_1>0$, corresponding to the same levels of the error metric; e.g., the contour of $A_{\varepsilon_1} = 10\%$ reaches slightly beyond $\Delta_1=1$ in Fig.~\ref{Fig: varepsilon_1 contours}c, whereas it only reaches extends up to somewhere near $\Delta_1 = -0.5$ in Fig.~\ref{Fig: varepsilon_1 contours}d. 
    \item Numerous sets of $\{\Delta_n,\Delta_1\}$ 
    can be suitably selected from the regions where all the metrics have low values. 
\end{enumerate}

\begin{figure}[h!]
  \centering
    \includegraphics[width=0.8\textwidth]{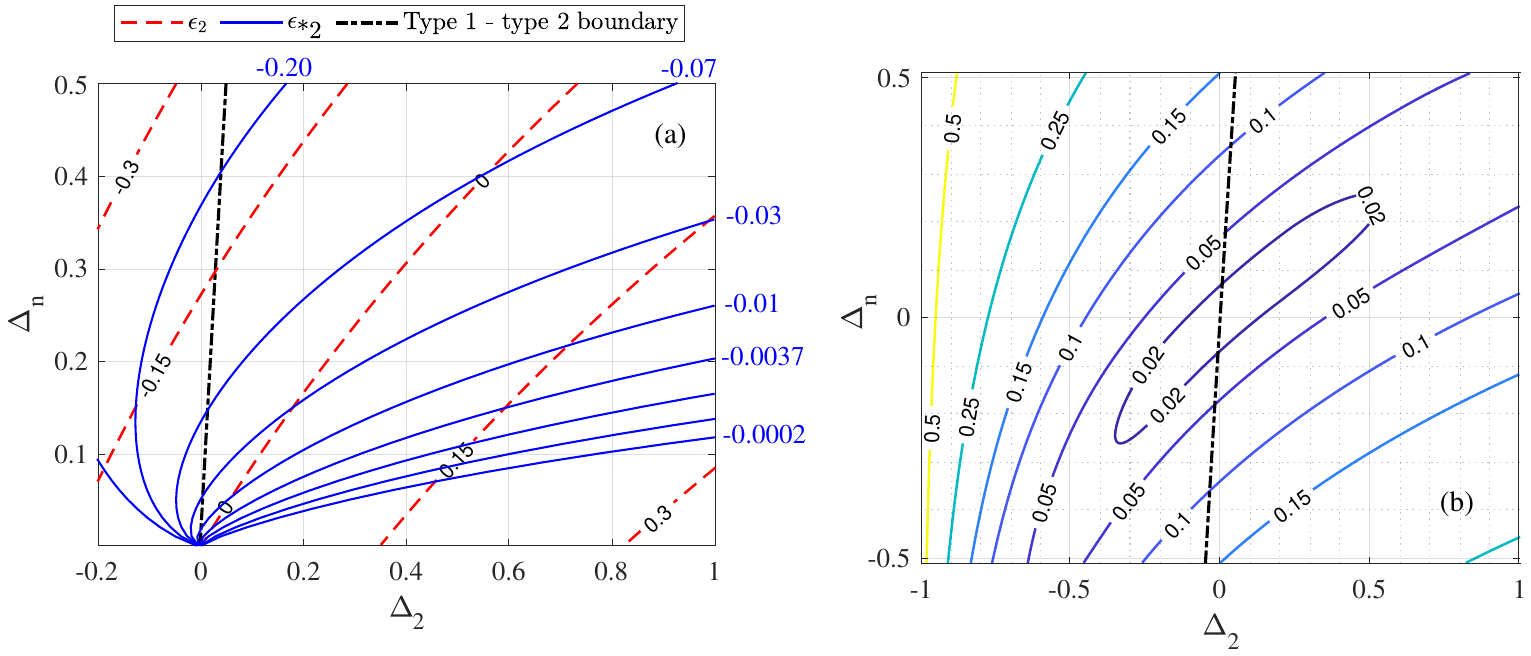}
  \caption{Contour plots of $\varepsilon_2(r)$ metrics for $\beta = 2, \gamma = 1$ and $n = 2$: (a) $\epsilon_2$ and $\epsilon_{*2}$, and (b) $A_{\varepsilon_2}$.}
  \label{Fig: varepsilon_2 contours}
\end{figure}

The contour plots of the metrics: $\epsilon_2$, $\epsilon_{*2}$ and $A_{\varepsilon_2}$, are obtained next for $\Delta_n\in\left(-0.5,0.5\right]$ and $\Delta_2 \in\left(-1,1\right]$, and shown in Fig.~\ref{Fig: varepsilon_2 contours}. The contours of $\epsilon_2$ and $\epsilon_{*2}$ are only shown for $\Delta_n>0$, as the contour patterns are very similar to those of $\varepsilon_1(r)$ metrics (Fig.~\ref{Fig: varepsilon_1 contours}). However, it can be seen from Eqs.~\ref{epsilon 1} and \ref{epsilon 2} that, for any set of $\left\{\Delta_n,\Delta_1,\Delta_2\right\}$ where $\Delta_1 = \Delta_2$, $\varepsilon_1(r)=-\kappa\varepsilon_2(r),\ \forall r\in\left[-r_{\text{max}},r_{\text{max}}\right]$. Therefore, for any such set of $\left\{\Delta_n,\Delta_1,\Delta_2\right\}$, from Eqs.~\ref{eps1 metrics} and \ref{eps2 metrics}, the values of $\epsilon_1$ and $\epsilon_{*1}$ are $-\kappa$ times that of $\epsilon_2$ and $\epsilon_{*2}$, respectively, and from Eqs.~\ref{eps1 area} and \ref{eps2 area}, the value of $A_{\varepsilon_1}$ is $\abs{\kappa}$ times that of $A_{\varepsilon_2}$. Hence, it is observed that the contours of $\varepsilon_1(r)$ metrics in Fig.~\ref{Fig: varepsilon_1 contours} are $\abs{\kappa}$ times denser ($\kappa = 3$ in this example) along the $\Delta_1$ axis, than the contours of the corresponding $\varepsilon_2(r)$ metrics in Fig.~\ref{Fig: varepsilon_2 contours}.

For obtaining a set of $\left\{\Delta_n,\Delta_1,\Delta_2\right\}$, which give low values of $\varepsilon_1(r)$ and $\varepsilon_2(r)$, first, a suitable set of $\left\{\Delta_n,\Delta_1\right\}$ is selected from the regions where all $\varepsilon_1(r)$ metrics have low values. Then, with the chosen value of $\Delta_n$, a suitable value of $\Delta_2$ is selected from the regions where all $\varepsilon_2(r)$ metrics have low values.   

\begin{figure}[h!]
  \centering
    \includegraphics[width=0.7\textwidth]{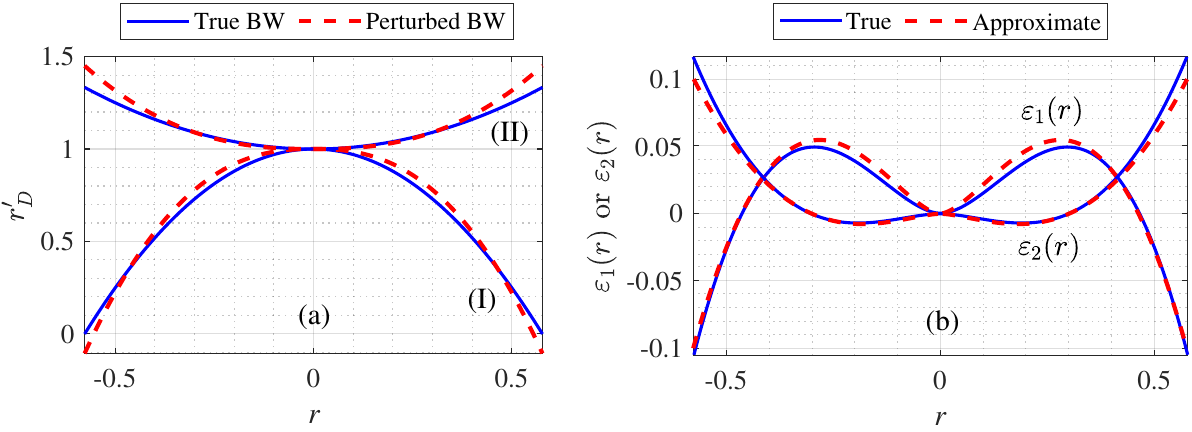}
  \caption{(a) $r^\prime_D$ for true $(\beta = 2, \gamma = 1, n = 2)$ and alternate $(\bar{\beta} = 3, \bar{\gamma} = 1.27, \bar{n} = 2.46)$ BW parameters, and (b) comparison of true $\varepsilon_1(r)$ and $\varepsilon_2(r)$ with their approximate expressions.}
  \label{Fig: true and alternate}
\end{figure}

For the considered numerical example, the change in $r^\prime_D$ for a typical set of $\left\{\Delta_n,\Delta_1,\Delta_2\right\} = \left\{0.23,0.42,0.73\right\}$, chosen suitably from Figs.~\ref{Fig: varepsilon_1 contours} and \ref{Fig: varepsilon_2 contours}, such that $\epsilon_1=-0.15$, $\epsilon_{*1}=0.084$, $A_{\varepsilon_1} = 0.054$, $\epsilon_2 = 0.15$, $\epsilon_{*2} = -0.012$ and $A_{\varepsilon_2} = 0.031$ are all small, is examined in details. The resulting differences between the true and alternate BW parameters are: $\Delta \beta = 1$, $\Delta \gamma = 0.27$ and $\Delta n = 0.46$, which are 50\%, 27\% and 23\% of their respective true values. $r^\prime_D(\textrm{I})$ and $r^\prime_D(\textrm{II})$ from Eq.~\ref{dr_dy} are shown for both the true and alternate BW parameters in Fig.~\ref{Fig: true and alternate}a. Despite the substantial differences, $r^\prime_D$ for both BW parameter sets are very similar to each other, which can lead to similar hysteretic force-deformation behaviours. Further, the difference between $r^\prime_D$ of the two BW parameter sets i.e., $\varepsilon_1(r)$ and $\varepsilon_2(r)$, and their approximate expressions given by Eqs.~\ref{epsilon 1} and \ref{epsilon 2}, respectively, are shown in Fig.~\ref{Fig: true and alternate}b; the approximations of both $\varepsilon_1(r)$ and $\varepsilon_2(r)$ match quite well with their true values $\forall r\in\left[-r_{\textrm{max}},r_{\textrm{max}}\right]$.

In this example, the alternate BW parameter sets may also lead to very similar dynamic responses, which is shown later in this work. This causes the BW parameters to be insensitive towards the dynamic responses of the structure. Therefore, when system identification is performed with measured dynamic responses from the structure, instead of the true parameters, the alternate parameters may be identified instead, indicating potential non-uniqueness of the BW parameters' estimates in system identification applications.

\section{Effect of the magnitude of $\left(\beta + \gamma\right)$ on alternate Bouc-Wen parameters}

From Eq.~\ref{eps1 metrics}, if the value of $\epsilon_1$ is fixed, the value of $r_{D,\textrm{avg}}^\prime$ from Eq.~\ref{rD_prime_area} can be used to obtain:
\begin{equation}{\label{Delta_1}}
    \left(1 + \Delta_1\right) = \left(\beta + \gamma\right)^{\Delta_n}e^{-\frac{n\epsilon_1}{n+1}}
\end{equation}
This value of $\left(1 + \Delta_1\right)$ is substituted in the expression of $\epsilon_{*1}$ from Eq.~\ref{eps1 metrics} to obtain:
\begin{equation}{\label{epsilon_*1}}
    \epsilon_{*1} = \frac{\Delta_n\left(n+1\right)\left(\beta + \gamma\right)}{ne\left(1 + \Delta_1\right)^{\frac{1}{\Delta_n}}} = \frac{\Delta_n\left(n+1\right)}{n}\exp{\frac{n\epsilon_1}{\Delta_n\left(n+1\right)} - 1}
\end{equation}
Similarly, the value of $\epsilon_2$ is fixed in Eq.~\ref{eps2 metrics}, the value of $r_{D,\textrm{avg}}^\prime$ is used from Eq.~\ref{rD_prime_area}, and the resulting value of $\left(1 + \Delta_2\right)$ is substituted in the expression of $\epsilon_{*2}$ from Eq.~\ref{eps2 metrics} to obtain:
\begin{equation}{\label{Delta_2 and epsilon_*2}}
     \left(1 + \Delta_2\right) = \left(\beta + \gamma\right)^{\Delta_n}e^{\frac{n\kappa\epsilon_2}{\left(n+1\right)}} \quad
     \text{and} \quad
     \epsilon_{*2} = -\frac{\left(n+1\right)\Delta_n}{n\kappa}\exp{-\left(1 + \frac{n\kappa\epsilon_2}{\left(n+1\right)\Delta_n}\right)}
\end{equation}
It can be observed from Eqs.~\ref{epsilon_*1} and \ref{Delta_2 and epsilon_*2} that $\epsilon_{*1}$ and $\epsilon_{*2}$ depend only on $\left\{\Delta_n,\epsilon_1\right\}$ and $\left\{\Delta_n,\epsilon_1,\kappa\right\}$, respectively. This means that for any value of $\left(\beta + \gamma\right)$, the contours of $\epsilon_1$ and $\epsilon_{*1}$ will always intersect at the same values of $\Delta_n$, and additionally, if the value of $\kappa$ is unchanged, $\epsilon_2$ and $\epsilon_{*2}$ will also intersect at the same values of $\Delta_n$. From Eqs.~\ref{Delta_1} and \ref{Delta_2 and epsilon_*2}, it is also noticed that both $\left(1+\Delta_1\right)$ and $\left(1+\Delta_2\right)$ are directly proportional to $\left(\beta + \gamma\right)^{\Delta_n}$. Therefore, for $\Delta_n>0$ and $\left(\beta+\gamma\right)>1$, the higher the magnitude of $\left(\beta+\gamma\right)$, the more the contours of $\varepsilon_1(r)$ and $\varepsilon_2(r)$ metrics will grow along $\Delta_1$ and $\Delta_2$ axes, respectively. Similarly, for $\Delta_n<0$ and $\left(\beta+\gamma\right)<1$, the lower the magnitude of $\left(\beta+\gamma\right)$, the more the contours of $\varepsilon_1(r)$ and $\varepsilon_2(r)$ metrics will grow along $\Delta_1$ and $\Delta_2$ axes, respectively. These observations are now illustrated with numerical examples.

\subsection{Numerical illustrations}{\label{Section: Examples beta gamma magnitude}}

\begin{figure}[h!]
  \centering
    \includegraphics[width=0.8\textwidth]{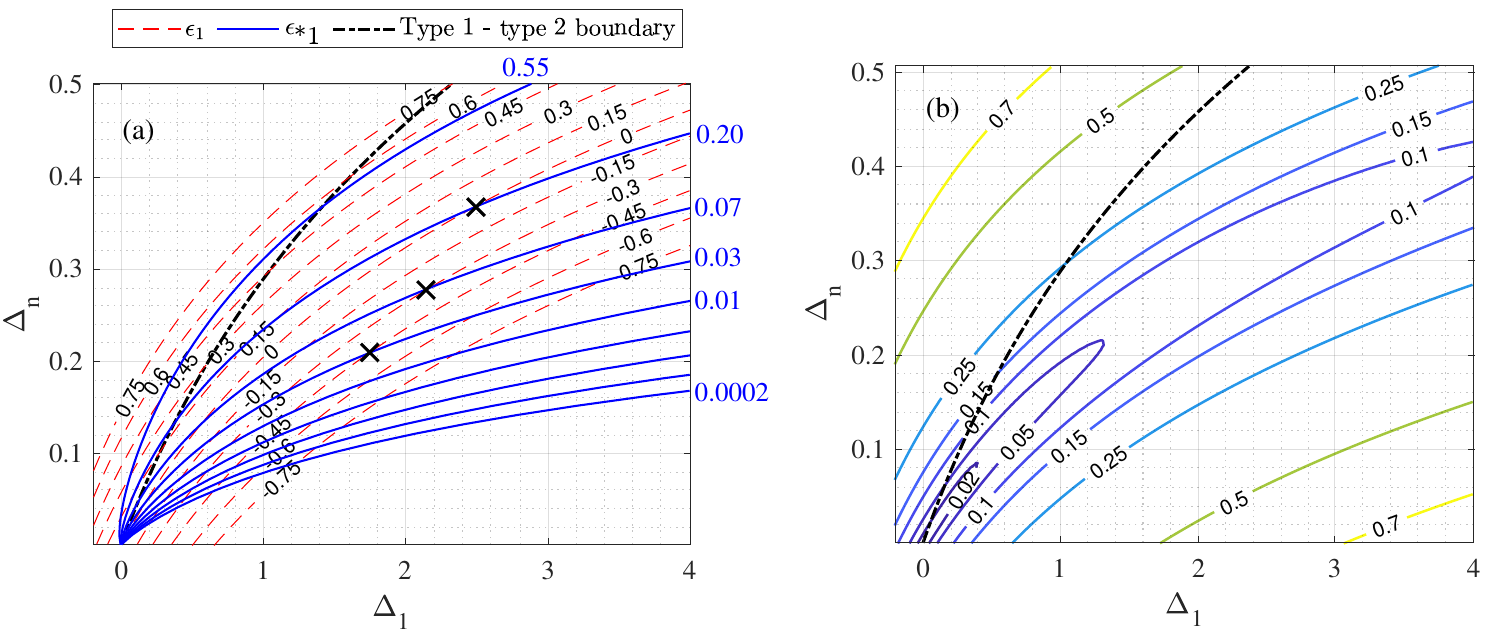}
  \caption{Contour plots of $\varepsilon_1(r)$ metrics for $\beta = 20, \gamma = 10$ and $n = 2$: (a) $\epsilon_1$ and $\epsilon_{*1}$, and (b) $A_{\varepsilon_1}$.}
  \label{Fig: varepsilon_1 contours beta 20}
\end{figure}

Considering $\beta = 20, \gamma = 10$ and $n = 2$, where $\beta + \gamma = 30$, the contours of $\varepsilon_1(r)$ metrics for $\Delta_n\in\left(0,0.5\right]$ and $\Delta_1 \in\left[-0.2,4\right]$, are shown in Fig.~\ref{Fig: varepsilon_1 contours beta 20}. Compared to the contours in Figs.~\ref{Fig: varepsilon_1 contours}a and c, where $\beta + \gamma = 3$, it can be seen that the contours in Fig.~\ref{Fig: varepsilon_1 contours beta 20} grow far more rapidly along the $\Delta_1$ axis, with low values of $\varepsilon_1(r)$ metrics also existing at relatively higher values of $\Delta_1$. However, the intersection points between the contours of $\epsilon_1$ and $\epsilon_{*1}$ have the same values of $\Delta_n$ in Figs.~\ref{Fig: varepsilon_1 contours}a and \ref{Fig: varepsilon_1 contours beta 20}a, as observed for the three black crosses in Fig.~\ref{Fig: varepsilon_1 contours beta 20}a. The intersection points between the type 1 - type 2 curves boundaries and the contours of $\epsilon_1$, $\epsilon_{*1}$ and  $A_{\varepsilon_1}$, and the tips of the $A_{\varepsilon_1}$ contours (especially $\forall A_{\varepsilon_1} = \left\{0.02,0.05\right\}$), can also be observed to have the same values of $\Delta_n$ in Figs.~\ref{Fig: varepsilon_1 contours} and \ref{Fig: varepsilon_1 contours beta 20}. As $\kappa$ remains the same for $\{\beta = 2,\gamma = 1,n = 2\}$ and $\{\beta = 20,\gamma = 10,n = 2\}$ i.e., $\kappa = 3$, these observations can also be seen with the contours of $\varepsilon_2(r)$ metrics. 

\begin{figure}[h!]
  \centering
    \includegraphics[width=0.8\textwidth]{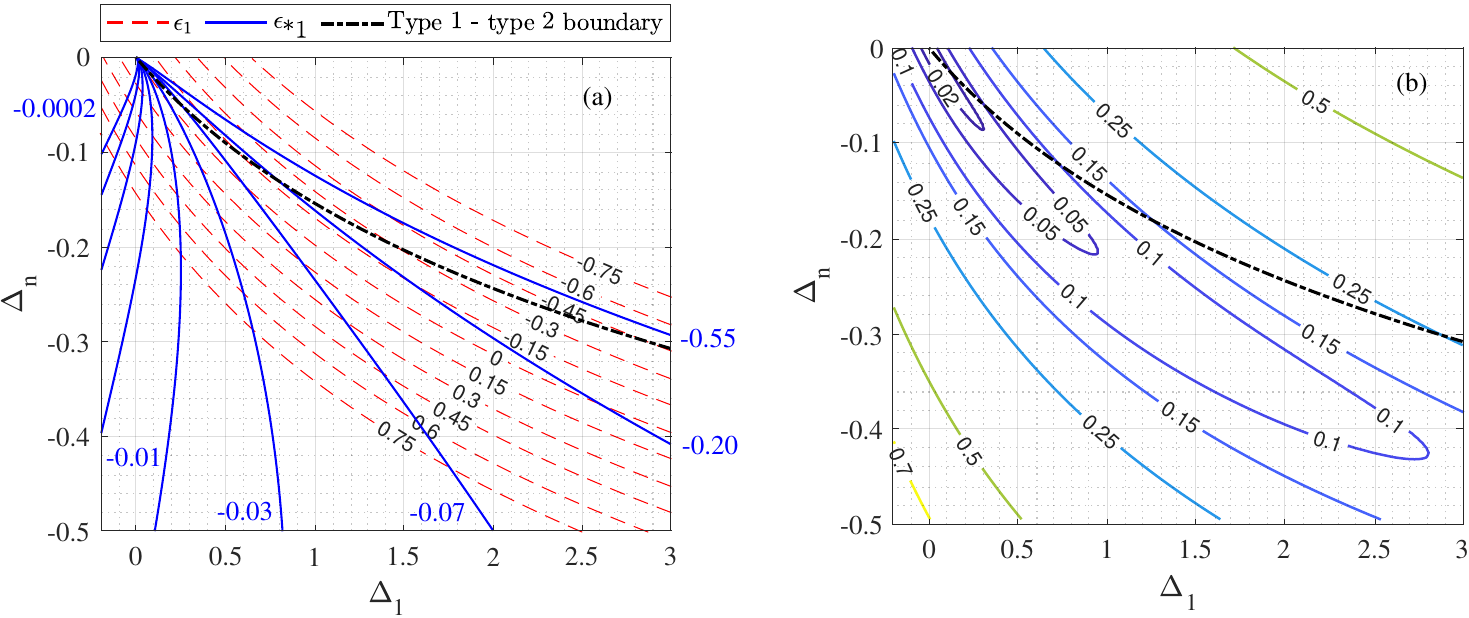}
  \caption{Contour plots of $\varepsilon_1(r)$ metrics for $\beta = 0.02, \gamma = 0.01$ and $n = 2$: (a) $\epsilon_1$ and $\epsilon_{*1}$, and (b) $A_{\varepsilon_1}$.}
  \label{Fig: varepsilon_1 contours beta 0p02}
\end{figure}

Now, considering $\beta = 0.02, \gamma = 0.01$ and $n = 2$, where $\beta + \gamma = 0.03$, the contours of $\varepsilon_1(r)$ metrics for $\Delta_n\in\left[-0.5,0\right)$ and $\Delta_1 \in\left[-0.2,3\right]$, are shown in Fig.~\ref{Fig: varepsilon_1 contours beta 0p02}. Similar to Fig.~\ref{Fig: varepsilon_1 contours beta 20}, the contours of $\varepsilon_1(r)$ metrics grow rapidly along the $\Delta_1$ axis, with low values of the metrics also existing at relatively higher values of $\Delta_1$. Also, for the intersection points between the contours of $\epsilon_1$ and $\epsilon_{*1}$, the intersection points between the type 1 - type 2 curves boundaries and the contours of $\epsilon_1$, $\epsilon_{*1}$ and  $A_{\varepsilon_1}$, and the tips of the $A_{\varepsilon_1}$ contours, all points have the the same values of $\Delta_n$ in Figs.~\ref{Fig: varepsilon_1 contours} and \ref{Fig: varepsilon_1 contours beta 0p02}. In this example as well, $\kappa=3$, and hence, these observations are valid for $\varepsilon_2(r)$ metrics.

In general, it was observed that, for high magnitudes of $\left(\beta+\gamma\right)$ as well as for magnitudes of  $\left(\beta+\gamma\right)\ll1$, low values of $\varepsilon_1(r)$ and $\varepsilon_2(r)$ metrics can be found at quite large values of $\Delta_1$ and $\Delta_2$, respectively, resulting in much larger magnitudes of $\Delta\beta$ and $\Delta\gamma$ which produce hysteretic force-deformation behaviours very similar to that of the true BW parameters. For example, considering Fig.~\ref{Fig: true and alternate}, the low values of error metrics with the true and alternate BW parameters: $\left\{\beta = 2, \gamma = 1, n = 2\right\}$ and $\left\{\bar{\beta} = 3, \bar{\gamma} = 1.27, \bar{n} = 2.46\right\}$, respectively, also occur with the true and alternate BW parameters: $\left\{\beta = 20, \gamma = 10, n = 2\right\}$ and $\left\{\bar{\beta} = 51.18, \bar{\gamma} = 21.58, \bar{n} = 2.46\right\}$, respectively. For the latter one with $\beta = 20$ and $\gamma = 10$, numerous such alternate sets of BW parameters can be found which deviate significantly from the true parameter values, and yet produce similar hysteretic behaviours.

\subsection{Equivalent Bouc-Wen systems}\label{Append}

The BW element governed by Eq.~\ref{BW evolution}, which is parameterized with the set $\left\{D_y,\beta,\gamma,n\right\}$, can be defined with multiple parameter sets, where $D_y>0$ and $n>1$. Let any two of the equivalent parameter sets be denoted as $\mathbb{S}_1=\left\{D_{y1},\beta_1,\gamma_1,n_1\right\}$ and $\mathbb{S}_2=\left\{D_{y2},\beta_2,\gamma_2,n_2\right\}$, respectively. The hysteretic deformations for $\mathbb{S}_1$ and $\mathbb{S}_2$ are denoted by $r_1$ and $r_2$, respectively, and the hysteretic forces by $f_{r1} = (1-\alpha)D_{y1}kr_1$ and $f_{r2} = (1-\alpha)D_{y2}kr_2$, respectively. For both sets to develop the same hysteretic force-deformation behaviours, $f_{r1} = f_{r2}$, which gives: 
\begin{equation}{\label{max force equating}}
    (1-\alpha)D_{y1}kr_1 = (1-\alpha)D_{y2}kr_2 \quad \implies \quad r_2 = \frac{D_{y1}}{D_{y2}}r_1
\end{equation}
The rates of change of the hysteretic force with the lumped mass displacement $y$ will be the same as well, i.e., $\dv{f_{r1}}{y}=\dv{f_{r2}}{y}$, which implies that:
\begin{equation}{\label{r_D equating}}
    (1-\alpha)kD_{y1}\dv{r_1}{y} = (1-\alpha)kD_{y2}\dv{r_2}{y} \quad \implies \quad r^\prime_{D1} = r^\prime_{D2}    
\end{equation}
where $r^\prime_{D1}$ and $r^\prime_{D2}$ are as defined in Eq.~\ref{dr_dy}. From Eq.~\ref{r_D equating}, $r^\prime_{D1} = r^\prime_{D2}$ is considered for both branches (I) and (II) to obtain:
\begin{equation}{\label{rD equivalent BW}}
    \frac{\beta_1 + \gamma_1}{\beta_2 + \gamma_2} = \frac{\abs{r_2}^{n_2}}{\abs{r_1}^{n_1}}
    \quad \textrm{and} \quad
    \frac{\beta_1 - \gamma_1}{\beta_2 - \gamma_2} = \frac{\abs{r_2}^{n_2}}{\abs{r_1}^{n_1}}
\end{equation}
Eq.~\ref{max force equating} is substituted in Eq.~\ref{rD equivalent BW} to obtain:
\begin{equation}{\label{rD equivalent BW 2}}
    \frac{\beta_1 + \gamma_1}{\beta_2 + \gamma_2} = \frac{\beta_1 - \gamma_1}{\beta_2 - \gamma_2} = \left(\frac{D_{y1}}{D_{y2}}\right)^{n_2}\abs{r_1}^{n_2 - n_1}    
\end{equation}
As all the parameters are constant, Eq.~\ref{rD equivalent BW 2} implies that $\abs{r_1}^{n_2 - n_1}$ must also be constant $\forall r_1$. This is possible $\forall r_1\in \mathbb{R} \setminus \{0\}$, if and only if $n_1 = n_2$. The case, where $r_1 = 0$, is neglected as, at $r_1 = 0$, $r_2 = f_{r1} = f_{r2} = \dv{f_{r1}}{y}=\dv{f_{r2}}{y} = 0,\ \forall n_1>1, n_2>1$. Therefore, the relationship between the two parameter sets is given as follows:
\begin{equation}{\label{BW equivalent}}
    n_1 = n_2 = n; \quad 
    \beta_1 = \left(\frac{D_{y1}}{D_{y2}}\right)^n\beta_2; \quad \textrm{and} \quad
    \gamma_1 = \left(\frac{D_{y1}}{D_{y2}}\right)^n\gamma_2
\end{equation}

Note that Ikhoune and Rodeller~\cite{ikhouane2007systems} have also shown this type of relationship between two BW models which belong to the same class, and for any input signal, produce exactly the same output, i.e., restoring force.

It can be seen from Eq.~\ref{rD equivalent BW 2} that, with the choice of a high value of $D_y$, the magnitude of $\left(\beta+\gamma\right)$ becomes high as well. This is often encountered in practice, where hysteretic systems with low yield displacement $D_y\ll1$ are expressed with a system considering $D_y = 1$, in order to remove $D_y$ from the governing equations altogether. However, this type of parameterization should be avoided, as substantially different alternate BW parameters producing very similar hysteretic behaviours will exist, which can result in significant erroneous identification of the BW parameters in inverse problems. Similarly, due to the same reason, the type of parameterization, which make $\left(\beta+\gamma\right)\ll1$ should also be avoided. In the example cases considered herein with $n = 2$, for one high magnitude of $\left(\beta + \gamma\right)$ i.e., $\beta = 20$ and $\gamma = 10$ (in Fig.~\ref{Fig: varepsilon_1 contours beta 20}) and for one magnitude of $\left(\beta + \gamma\right)\ll 1$ i.e., $\beta = 0.02$ and $\gamma = 0.01$ (in Fig.~\ref{Fig: varepsilon_1 contours beta 0p02}), it is shown that significantly different alternate BW parameters can exist. 





\section{Similar responses with alternate Bouc-Wen parameters}\label{Section: Similar responses}

\begin{figure}[h!]
  \centering
  \includegraphics[width=0.9\textwidth]{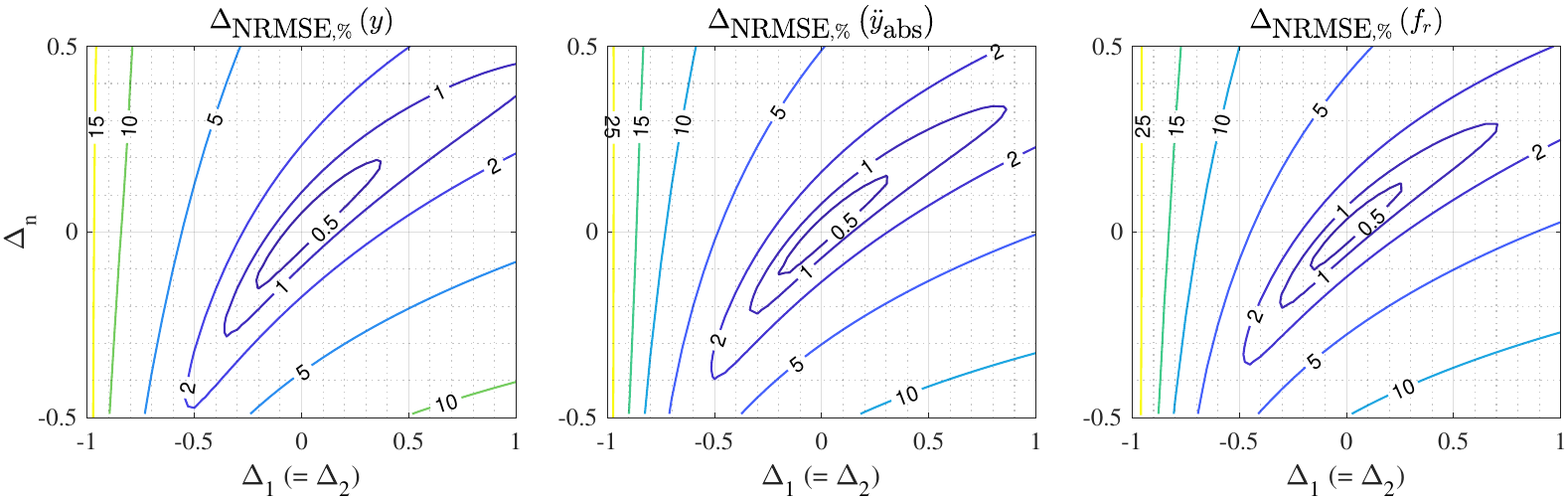}
  \caption{$\Delta_{\textrm{NRMSE,\%}}(\cdot)$ of displacement $y$, absolute acceleration $\ddot{y}_\textrm{acc}$ and hysteretic restoring force $f_r$, for alternate BW parameters, with El Centro ground motion.}
  \label{Fig: El Centro contours delta2 = delta1}
\end{figure}

In order to validate the similarity of hysteretic force behaviours with the alternate BW parameters, the SDOF lumped mass hysteretic system described by Eqs.~\ref{SDOF EOM} and~\ref{BW evolution} is considered, when subjected to the El Centro ground motion, with the following properties: $m = 1$ kg, $c = 0.5$ Ns/m, $\alpha = 0.1$, $k = 100$ N/m, $D_y = 0.0365$ m, $\beta = 2, \gamma = 1$ and $n = 2$. The dynamic responses of the system are computed with the fourth-order Runge-Kutta method with a time step of 0.001 s. Dynamic responses are also computed for alternate sets of BW parameters, which are obtained by varying the original parameters considering $\Delta_n\in\left(-0.5,0.5\right]$ and $\Delta_1 \in\left(-1,1\right]$. Also, $\Delta_2 = \Delta_1$ is considered, which implies $\frac{\Delta\beta + \Delta\gamma}{\beta + \gamma} = \frac{\Delta\beta - \Delta\gamma}{\beta - \gamma} \implies \Delta\beta/\beta = \Delta\gamma/\gamma = \Delta_1$. The Normalized Root Mean Squared Error (NRMSE) percentages $\Delta_{\textrm{NRMSE,\%}}(\cdot)$ are computed for the obtained dynamic responses with respect to the true responses obtained with the actual (true) BW parameters. $\Delta_{\textrm{NRMSE,\%}}(x)$ for a response quantity $\bar{x}$ obtained with any set of alternate BW parameters $\left\{\bar{\beta},\bar{\gamma},\bar{n}\right\}$, with respect to the corresponding true response $x$, is defined as: 
\begin{equation}{\label{NRMSE}}
    \Delta_{\textrm{NRMSE,\%}}(x) = \frac{\frac{1}{L} \sqrt{\sum\limits_{k=1}^{L}\left(\bar{x}_k-x_k\right)^2}}{\textrm{max}\left(x\right)-\textrm{min}\left(x\right)}\times{100\%}
\end{equation}
where $\bar{x}_k$ and $x_k$ are the values of $\bar{x}$ and $x$ at the $k$th time instant, respectively, and $L$ is the total number of time instants (length of data). The values of $\Delta_{\textrm{NRMSE,\%}}(\cdot)$ thus obtained for the displacement $y$ and absolute acceleration $\ddot{y}_\textrm{abs}$ of the lumped mass, and the hysteretic restoring force $f_r$ in the Bouc-Wen element, are shown in Fig.~\ref{Fig: El Centro contours delta2 = delta1}. It can be seen that there is a significant region of $\left\{\Delta_n,\Delta_1,\Delta_2\right\}$ with very low values of $\Delta_{\textrm{NRMSE,\%}}(f_r)$ (such as <1\% error), indicating that the alternate BW parameters from this region result in extremely similar time histories of the hysteretic force. Therefore, these~alternate BW parameters also lead to very low values of NRMSE errors for dynamic responses, as observed for $y$ and $\ddot{y}_\textrm{abs}$ in Fig.~\ref{Fig: El Centro contours delta2 = delta1}. It is to be noted that these regions of low $\Delta_{\textrm{NRMSE,\%}}(\cdot)$ also correspond to the regions of low values of $\varepsilon_1(r)$ and $\varepsilon_2(r)$ metrics, shown in Figs.~\ref{Fig: varepsilon_1 contours} and \ref{Fig: varepsilon_2 contours}, respectively.

\begin{figure}[h!]
  \centering
  \includegraphics[width=0.95\textwidth]{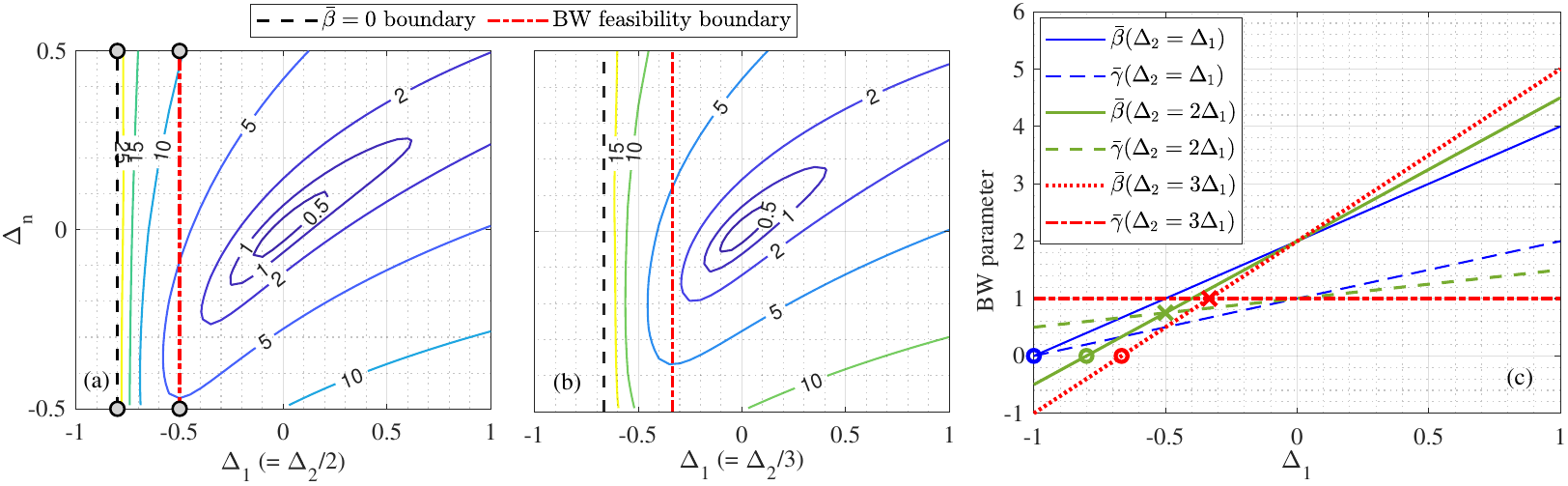}
  \caption{(a-b) $\Delta_{\textrm{NRMSE,\%}}(f_r)$ with El Centro ground motion, for $\Delta_2 = 2\Delta_1$ and $\Delta_2 = 3\Delta_1$, respectively, and (c) variation of $\bar{\beta}$ and $\bar{\gamma}$ with $\Delta_1$, for various relationships between $\Delta_1$ and $\Delta_2$.}
  \label{Fig: El Centro contours diff combos}
\end{figure}

In Figs.~\ref{Fig: varepsilon_1 contours} and \ref{Fig: varepsilon_2 contours}, it was observed for $\beta = 2$, $\gamma = 1$ and $n=2$ that the magnitudes of $\varepsilon_2(r)$ metrics increase slower with $\Delta_2$, than the magnitudes of $\varepsilon_1(r)$ metrics do with increase in $\Delta_1$. Therefore, as larger values of $\Delta_2$ can be considered, apart from considering $\Delta_2 = \Delta_1$, $\Delta_2 = 2\Delta_1$ and $\Delta_2 = 3\Delta_1$ are also considered here for the same ranges of $\Delta_n$ and $\Delta_1$. The resulting contours of $\Delta_{\textrm{NRMSE,\%}}(f_r)$ under the El Centro ground motion for $\Delta_2 = 2\Delta_1$ and $\Delta_2 = 3\Delta_1$ are shown in Figs.~\ref{Fig: El Centro contours diff combos}a and b, respectively. The variation of $\bar{\beta}$ and $\bar{\gamma}$ with $\Delta_1$, for $\Delta_2 = \Delta_1$, $\Delta_2 = 2\Delta_1$ and $\Delta_2 = 3\Delta_1$ are also compared in Fig.~\ref{Fig: El Centro contours diff combos}c. From Figs.~\ref{Fig: El Centro contours diff combos}a and b, several combinations of alternate BW parameters $\bar{\beta}$, $\bar{\gamma}$ and $\bar{n}$ can be obtained from the regions where $\Delta_{\textrm{NRMSE,\%}}(f_r)$ is very low, which result in very similar time histories of hysteretic forces and dynamic responses, and can be different than those obtained considering $\Delta_2 = \Delta_1$ from Fig.~\ref{Fig: El Centro contours delta2 = delta1}. Other such combinations of alternate BW parameters can be found by considering other relationships between $\Delta_1$ and $\Delta_2$.

However, from Fig.~\ref{Fig: El Centro contours diff combos}c, for the $\Delta_2 = 2\Delta_1$ and $\Delta_2 = 3\Delta_1$ cases, it can be seen that $\beta$ becomes zero in the considered range of $\Delta_1\in\left(-1,1\right]$, which are shown by circular markers. As the system response is unstable for $\bar{\beta}<0$~\cite{ikhouane2007dynamic}, it can be seen in Figs.~\ref{Fig: El Centro contours diff combos}a and b that all contours exist in the region for $\bar{\beta}>0$ only.
In Fig.~\ref{Fig: El Centro contours diff combos}c, for $\Delta_2 = 2\Delta_1$ and $\Delta_2 = 3\Delta_1$, it can also be observed that the feasibility condition of the BW parameters $-\bar{\beta}\leq\bar{\gamma}\leq\bar{\beta}$ is violated in the regions left of the respective intersection points of $\bar{\beta}$ and $\bar{\gamma}$ lines, which are marked with crosses. The respective boundaries of feasible BW parameters are also shown in Figs.~\ref{Fig: El Centro contours diff combos}a and b. In both figures, the regions of $\left\{\Delta_n,\Delta_1,\Delta_2\right\}$ having low values of $\Delta_{\textrm{NRMSE,\%}}(f_r)$, such as $\Delta_{\textrm{NRMSE,\%}}(f_r)<2 \%$,  where suitable alternate BW parameters can be found, mostly lie in the respective regions of feasible BW parameters.

\begin{figure}[h!]
  \centering
  \includegraphics[width=0.75\textwidth]{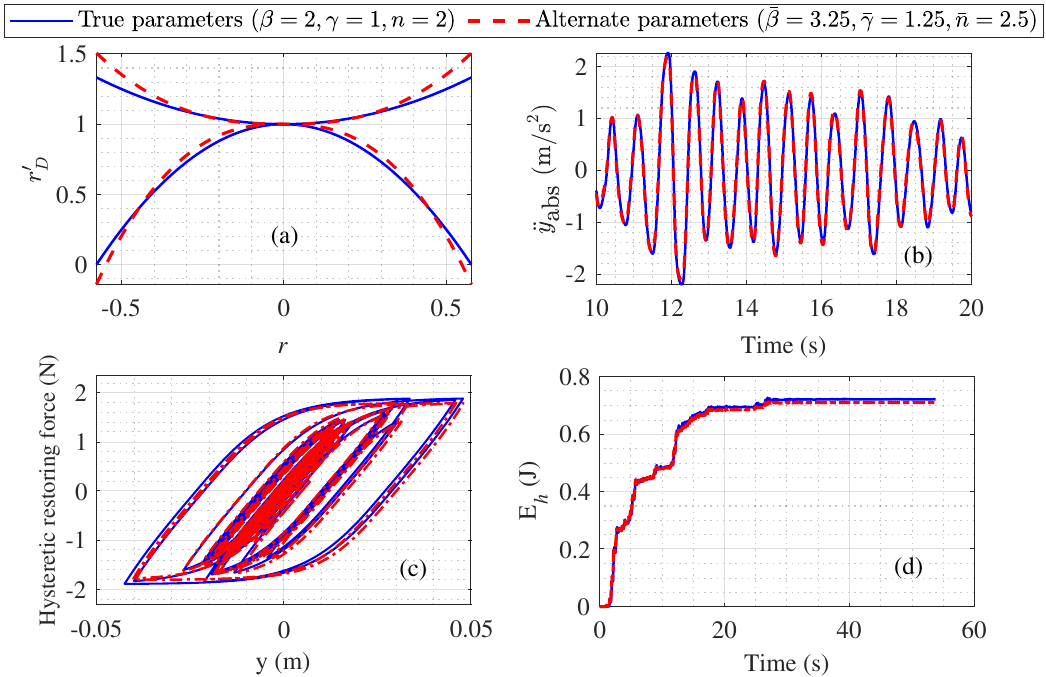}
  \caption{Typical example of true and alternate BW parameters showing similar hysteretic restoring force behaviour and dynamic responses, with El Centro ground motion.}
  \label{Fig: El Centro typical examples}
\end{figure}

The behaviour of the considered SDOF hysteretic system under the excitation of El Centro ground motion, with a typical set of alternate BW parameters from Fig.~\ref{Fig: El Centro contours diff combos}a, is examined in detail. The chosen set of $\left\{\Delta_n,\Delta_1,\Delta_2\right\}$ is $\left\{0.25,0.5,1\right\}$, where $\Delta_{\textrm{NRMSE,\%}}(f_r) = 0.9\ \%$, $\Delta_{\textrm{NRMSE,\%}}(y) = 0.6\ \%$ and $\Delta_{\textrm{NRMSE,\%}}(\ddot{y}_\textrm{acc}) = 0.8\ \%$. Also, the corresponding $\varepsilon_1(r)$ and $\varepsilon_2(r)$~metrics are: $\epsilon_1 = -0.205$, $\epsilon_{*1}=0.078$, $A_{\varepsilon_1}=0.058$, $\epsilon_2=0.212$, $\epsilon_{*2}=-0.008$ and $A_{\varepsilon_2}=0.048$. The resulting alternate BW parameters are $\bar{\beta} = 3.25, \bar{\gamma} = 1.25$ and $\bar{n} = 2.5$. The absolute acceleration $\ddot{y}_\textrm{acc}$, the hysteretic force-deformation behaviour i.e., hysteretic restoring force $f_r$ versus lumped mass displacement $y$, and the hysteretic energy dissipated ($E_h$ in Eq.~\ref{hyst_ener}), obtained with the alternate BW parameters, are compared with those obtained for the true BW parameters in Fig.~\ref{Fig: El Centro typical examples}. A comparison between the scaled rate of change of the hysteretic force with displacement $r^\prime_D = D_y\dv*{r}{y}$ (Eq.~\ref{dr_dy}) for the two BW parameter sets is also shown in Fig.~\ref{Fig: El Centro typical examples}. It can be clearly observed that the alternate BW parameters result in hysteretic behaviour and dynamic responses, which are very similar to those obtained with the true BW parameters, despite the alternate BW parameters deviating substantially from the true BW parameters. 

When reinforced concrete (RC) structures are subjected to earthquakes, they exhibit hysteretic force-deformation behaviour and undergo damage. For quantifying the damage imparted to the RC members, several damage indices have been developed which depend on the maximum deformation and/or the total hysteretic energy dissipated~\cite{powell1988seismic,park1985mechanistic,cosenza1993use,bozorgnia2003damage,colombo2005damage}. The Park-Ang damage index~\cite{park1985mechanistic} is one such popular index which is expressed as a linear combination of the maximum deformation $y_{\textrm{max}}$ and the total hysteretic energy dissipated $E_{h,\textrm{tot}}$ as:
\begin{equation}{\label{Park_Ang}}
    DI = \frac{y_{\textrm{max}}}{y_{\textrm{ult}}} + \frac{\delta_E}{F_{y}y_{\textrm{ult}}}E_{h,\textrm{tot}}
\end{equation}
where $y_{\textrm{ult}}$ and $F_{y} = kD_y$ are the ultimate deformation capacity and the yield strength, respectively, and $\delta_E$ is a weighting factor obtained empirically from past data. The Park-Ang index has been calibrated for the damage states: slight for $DI<0.2$, moderate for $0.2<DI<0.5$, severe for $0.5<DI<1$ and collapse for $DI>1$. For an RC structure being modelled with the considered SDOF BW hysteretic system and subjected to the El Centro ground motion, the response and the hysteretic behaviours with the true $\left(\beta = 2, \gamma = 1, n = 2\right)$ and alternate $\left(\bar{\beta} = 3.25, \bar{\gamma} = 1.25, \bar{n} = 2.5\right)$ parameter sets, are the same as that shown in Fig.~\ref{Fig: El Centro typical examples}. Assuming $y_{\textrm{ult}} = 6D_y$ and $\delta_E = 0.10$, the damage indices considering the true and alternate BW parameters are 0.299 and 0.298, respectively. The similarity in these values is expected and can be observed with all such alternate parameter sets that result in similar hysteretic behaviours. This suggests that, for structural health monitoring purposes, it may be more judicious to directly estimate the damage index as a measure of the damage suffered by the structure under ground motion excitations, instead of attempting to estimate the BW parameters.


\subsection{Effect of input excitation on alternate parameters}{\label{Section: effect of input}}

\begin{figure}[h!]
  \centering
  \includegraphics[width=0.6\textwidth]{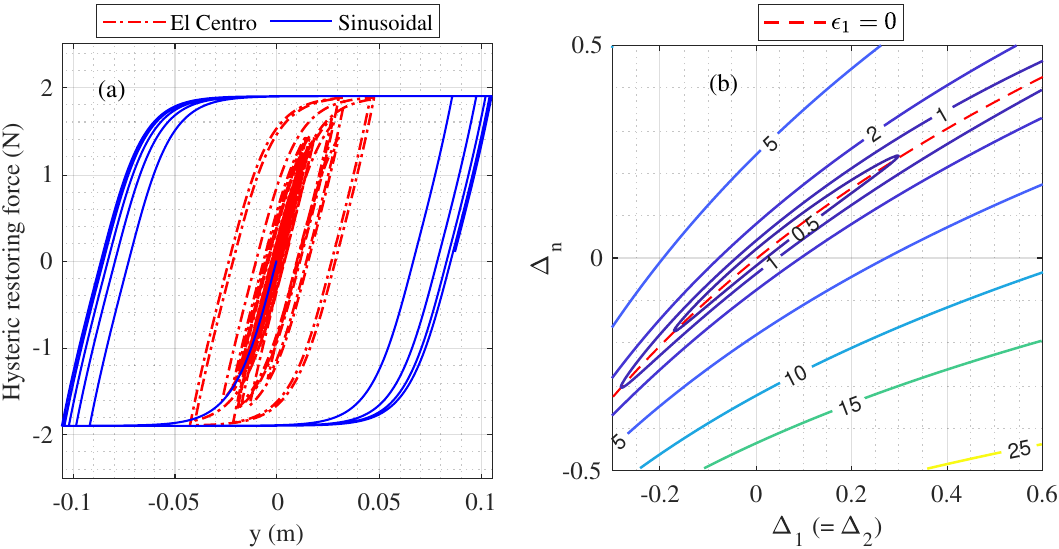}
  \caption{(a) Comparison of hysteretic behaviours obtained with El Centro ground motion and sinusoidal excitation, and (b) $\Delta_{\textrm{NRMSE,\%}}(f_r)$ contours considering  $\Delta_2 = 2\Delta_1$.}
  \label{Fig: Sine 2p5}
\end{figure}

For the considered SDOF system, the effect of alternate BW parameters is also examined under the sinusoidal ground excitation $f(t) = -2.5\sin{\pi t}$ from $t=0$ s to $t = 10$~s. The hysteretic behaviour $f_r$ versus $y$, with the true BW parameters, is shown in Fig.~\ref{Fig: Sine 2p5}a, and also compared with the hysteretic behaviour obtained previously with the El Centro ground motion. In this case, $y$ is much larger and the hysteretic loop is much fatter than the case considering El Centro ground motion, with the hysteretic displacement $r$ reaching $\pm r_{\textrm{max}}$ and sustaining it for a considerable amount of time during every loop of the sinusoidal force. Therefore, in order to produce similar hysteretic behaviour, alternate BW parameters should also result in their respective hysteretic forces reaching from low values to their maximum near $r=\pm r_{\textrm{max}}$. Now, this part of the hysteresis is described by branch (I)(Eqs.~\ref{BW branches}-\ref{dr_dy}), with the scaled rate of change of the hysteretic force with displacement in branch (I) for the true BW parameters $r^\prime_D (I) = 0$ at $r=\pm r_{\textrm{max}}$. Hence, the scaled rate of change of the hysteretic force with displacement (Eq.~\ref{dr_dy}) in branch (I) for the alternate BW parameters $\bar{r}^\prime_D (I)$ should reach zero near $r=\pm r_{\textrm{max}}$. As the metric $\epsilon_1 = \frac{\varepsilon_1(r_{\text{max}})}{r_{D,\textrm{avg}}^\prime}$ (Eq.~\ref{eps1 metrics}) is a measure of the deviation of $\bar{r}^\prime_D (I)$ from $r^\prime_D (I)$ at $r=\pm r_{\textrm{max}}$, it should be close to zero for the alternate BW parameters.

Note that the alternate BW parameters, whose hysteretic forces reach exactly their maximum at $r=\pm r_{\textrm{max}}$, satisfy: 
\begin{equation}{\label{rmax_same}}
    \left(\bar{\beta} + \bar{\gamma}\right)^{-\frac{1}{n + \Delta n}} = \left(\beta + \gamma\right)^{-\frac{1}{n}}
\end{equation} 
This is also obtained when either the expression for $\epsilon_1=0$ or the expression for the non-approximated form of $\varepsilon_1(r_{\text{max}})$ is algebraically rearranged. As the alternate parameters producing similar hysteretic behaviour need $\epsilon_1$ to be close to zero, this means that the alternate parameters closely follow the relationship in Eq.~\ref{rmax_same} as well.

\begin{figure}[h!]
  \centering
  \includegraphics[width=0.9\textwidth]{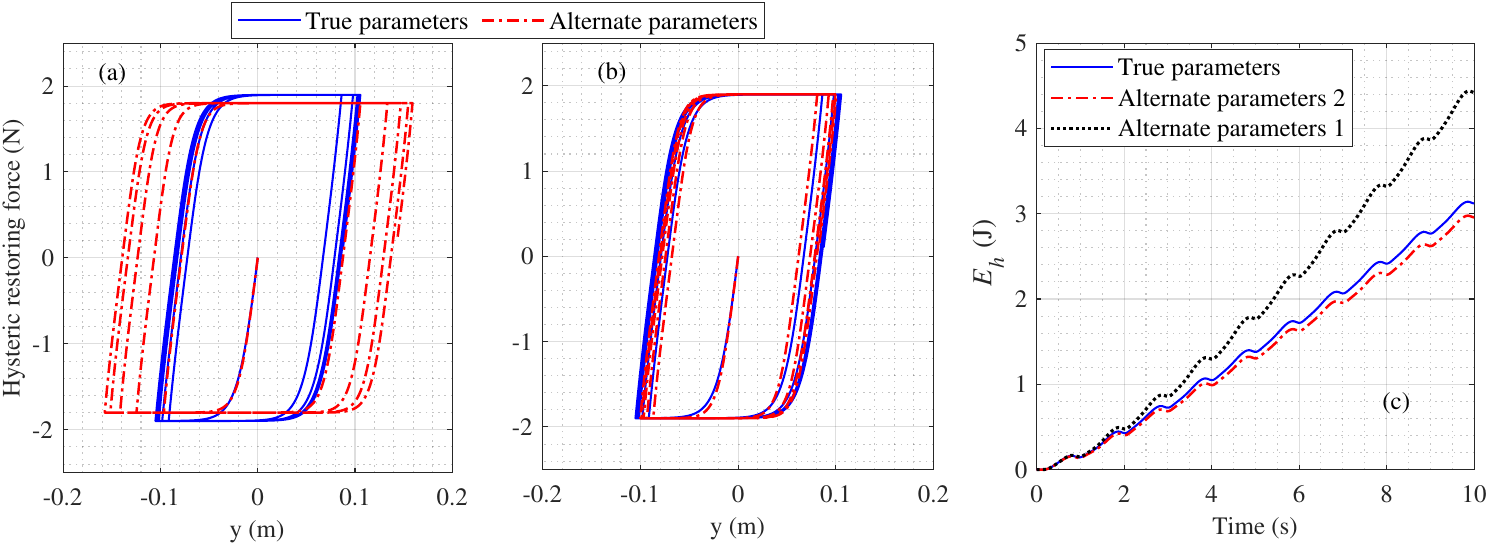}
  \caption{For sinusoidal excitation, (a-b) comparison of hysteretic behaviour between true parameters and alternate parameters (a) set 1: $(\bar{\beta} = 3.25, \bar{\gamma} = 1.25, \bar{n} = 2.5)$ and (b) set 2: $(\bar{\beta} = 3.25, \bar{\gamma} = 1.25, \bar{n} = 2.74)$; and (c) comparison of hysteretic energy dissipation between true parameters and alternate parameter sets 1 and 2.}
  \label{Fig: Sine 2p5 examples}
\end{figure}

When hysteretic forces and dynamic responses are obtained with different sets of $\left\{\Delta_1,\Delta_2,\Delta_n\right\}$, it can be seen that the NRMSE errors are very low only near $\epsilon_1 = 0$. This is observed with the contours of $\Delta_{\textrm{NRMSE,\%}}(f_r)$ in Fig.~\ref{Fig: Sine 2p5}b, which are obtained for $\Delta_1\in[-0.3,0.6]$ and $\Delta_n\in\left(-0.5,0.5\right]$, while considering $\Delta_2 = 2\Delta_1$. Compared to the values of $\Delta_{\textrm{NRMSE,\%}}(f_r)$ obtained with the El Centro ground motion in Figs.~\ref{Fig: El Centro contours delta2 = delta1}, \ref{Fig: El Centro contours diff combos}a and b, the regions for low values of $\Delta_{\textrm{NRMSE,\%}}(f_r)$ are very narrow in Fig.~\ref{Fig: Sine 2p5}b. 

In Fig.~\ref{Fig: Sine 2p5 examples}a, the hysteretic force-deformation behaviour with the alternate BW parameters, considered in Fig.~\ref{Fig: El Centro typical examples}, is compared with the true behaviour. Unlike Fig.~\ref{Fig: El Centro typical examples}c, the hysteretic behaviours are quite different for the true and alternate parameters in this case, as $\epsilon_1 = -0.205$ is not close to zero for this set of alternate parameters. In fact, in this case, it's $\bar{r}^\prime_D (I)$ becomes zero slightly earlier than $|r|= r_{\textrm{max}}$, as seen in Fig.~\ref{Fig: El Centro typical examples}a, which makes the produced hysteresis loops even fatter. Another set of BW parameters is chosen, obtained with the set $\left\{\Delta_1,\Delta_2,\Delta_n\right\} = \left\{0.5,1,0.37\right\}$, which is a point in Fig.~\ref{Fig: Sine 2p5}b near the line $\epsilon_1 = 0$. The remaining $\varepsilon_1(r)$ and $\varepsilon_2(r)$ metrics are low for this point and the resulting alternate BW parameters $\bar{\beta} = 3.25, \bar{\gamma} = 1.25$ and $\bar{n} = 2.74$ produce hysteretic behaviour very similar to the true behaviour, as observed in Fig.~\ref{Fig: Sine 2p5 examples}b. Referring to the alternate parameter sets of Figs.~\ref{Fig: Sine 2p5 examples}a and b as sets 1 and 2, respectively, the corresponding hysteretic energy dissipations ($E_h$ in Eq.~\ref{hyst_ener}) under the sinusoidal excitation are compared with that for the true parameters in Fig.~\ref{Fig: Sine 2p5 examples}c. As expected, only the $E_h$ for set 2 matches closely with that from the true parameters. Further, the Park-Ang Damage indices $DI$ (Eq.~\ref{Park_Ang}) for sets 1 and 2 are 1.224 and 0.792, respectively. As $DI = 0.832$ for the true parameters, only the damage estimation with set 2 matches well.

It can be concluded that the alternate BW parameters are dependent on the extent of the hysteresis action being developed by the input excitation. For lower levels of excitations, the regions of alternate BW parameters tend to be relatively larger than the regions corresponding to higher levels of excitations. Nonetheless, the insensitivity-induced potential non-uniqueness in the system identification of the BW parameters may still exist even for high levels of excitations, which cause significant development of the hysteresis action.

\subsection{Similarity of inelastic displacement ratio spectra}{\label{Section: Cr spectrum}}

In this section, the similarities between the inelastic displacement ratio spectra of the considered hysteretic SDOF systems, obtained with the true and alternate BW parameters, are illustrated. For an SDOF system subjected to a ground motion and having a time period $T$, its inelastic displacement ratio $C_R (T)$, computed considering a constant yield strength reduction factor $R$, is defined as:
\begin{equation}
    C_R(T) = \frac{y_{\textrm{max},R}}{y_{{\textrm{max},R|}_{R=1}}}
\end{equation}
where $y_{{\textrm{max},R|}_{R=1}}$ is the peak elastic displacement of the corresponding linear SDOF system ($\alpha = 1$ in Eq~\ref{SDOF EOM}) and $y_{\textrm{max},R}$ is the peak inelastic displacement of the hysteretic SDOF system. The hysteretic system has a yield strength of $1/R$ times the minimum required by the system to stay in the elastic zone under the applied ground motion, which results in its yield displacement being computed as $D_y = y_{{\textrm{max},R|}_{R=1}}/R$.

\begin{figure}[h!]
  \centering
  \includegraphics[width=1\textwidth]{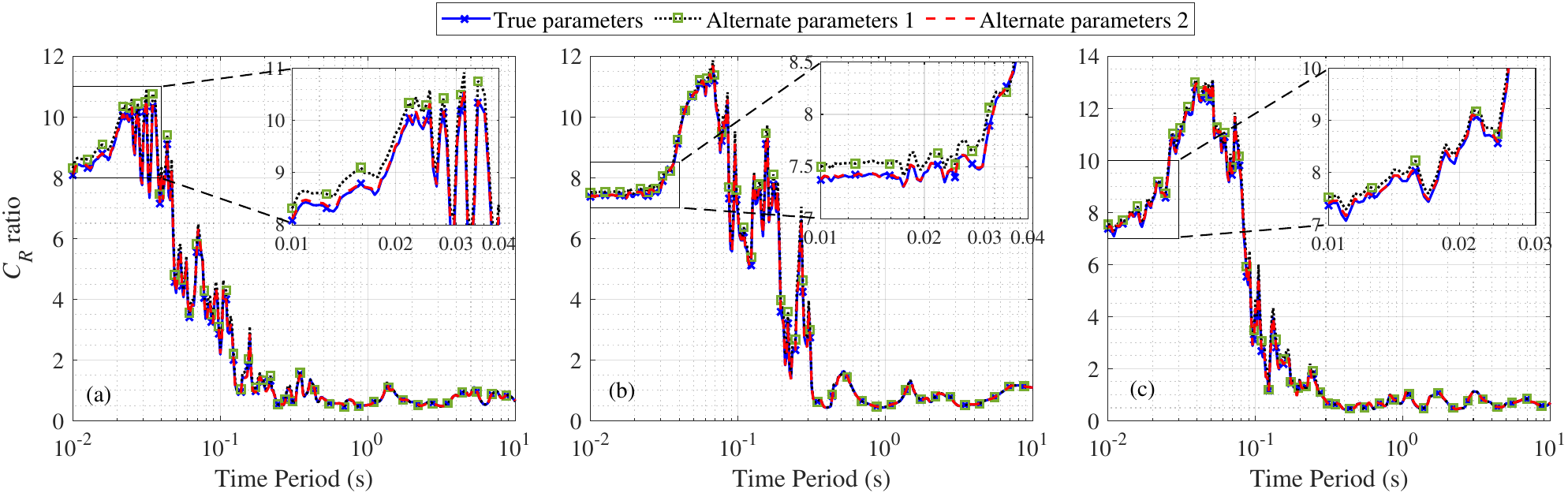}
  \caption{$C_R$ spectra obtained with true parameters $\left(\beta = 2, \gamma = 1, n = 2\right)$ and alternate parameters (set 1: $\left(\bar{\beta} = 3.25, \bar{\gamma} = 1.25, \bar{n} = 2.5\right)$, set 2: $\left(\bar{\beta} = 3.25, \bar{\gamma} = 1.25, \bar{n} = 2.74\right)$) for (a) El Centro, (b) Kobe and (c) Northridge ground motions}
  \label{Fig: C_R spectra}
\end{figure}

The $C_R (T)$ spectra are evaluated for three ground motion time histories from three earthquakes: El Centro 1940, Northridge 1994 and Kobe 1995. The three considered ground motions have peak ground accelerations of 0.35 g, 0.45 g and 0.6 g, respectively. A damping ratio of $2\%$ and $R=2$ are considered. The $C_R (T)$ spectra are evaluated for the true parameters $\beta = 2, \gamma = 1$ and $n = 2$, as well as the alternate parameter sets 1 and 2 considered in Fig.~\ref{Fig: El Centro typical examples}. The computed $C_R$ spectra are shown in Fig.~\ref{Fig: C_R spectra}. It is observed that the $C_R (T)$ ratios are high for lower values of the time period $T$. This can be attributed to the fact that the extent of hysteresis action is high, and very fat hysteresis loops are formed with low values of $T$, which results in the inelastic deformation being much higher than the elastic deformation of the corresponding linear system. 

It can be seen that the $C_R (T)$ spectra of both sets of alternate parameters match very well that of the true parameters. However, the spectrum is slightly higher for the set 1 parameters, at low values of $T$ where the hysteresis loops are quite fat. This is due to the alternate parameters of set 1 being unable to give similar hysteretic behaviours and producing even fatter hysteresis loops in such cases, as discussed in the previous Section~\ref{Section: effect of input} and observed in Fig.~\ref{Fig: El Centro typical examples}a. With fatter hysteresis loops being produced by set 1, the inelastic displacements are higher than those with the true parameters (or set 2 of the alternate parameters), and therefore, the corresponding $C_R (T)$ ratios are higher as well. Nonetheless, the difference is very minimal, and overall, the $C_R (T)$ spectra for both the alternate parameter sets can be said to be very similar to that for the true parameters.

\section{Erronous system identification of the Bouc-Wen parameters}\label{Section: Erronous system ID}

\begin{figure}[h!]
  \centering
  \includegraphics[width=0.7\textwidth]{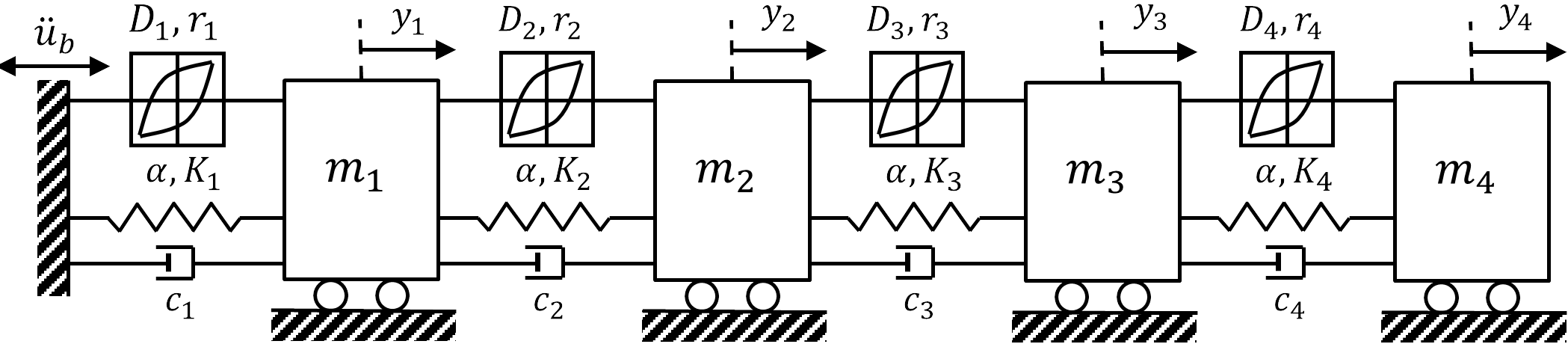}
  \caption{Four DOFs model with dissipative Bouc-Wen restoring force mechanisms}
  \label{Fig: four_DOF_BW_model}
\end{figure}

In this section, the potential non-uniqueness in the system identification of the BW parameters in hysteretic structures is illustrated with a numerical example. A four DOFs mass-spring-dashpot chain-type model with BW restoring force elements, subjected to base excitation $\Ddot{u}_b\left(t\right)$, is considered, as shown in Fig.~\ref{Fig: four_DOF_BW_model}. The governing equation of motion of the system is
\begin{equation}{\label{fourDOF_EOM}}
    \mathbf{M}\ddot{\mathbf{y}} + \mathbf{C}\dot{\mathbf{y}} + \alpha\mathbf{K}_{\textrm{lin}}\mathbf{y} + \left(1-\alpha\right)\mathbf{K}\mathbf{r} = -\mathbf{M}\mathbf{I}_b \ddot{u}_b\left(t\right)
\end{equation}
where $\mathbf{y}=\left( y_1,y_2,y_3,y_4\right)^T$ is the displacement vector (relative to base), $\mathbf{r}=\left( r_1 ,r_2 ,r_3 ,r_4 \right)^T$ is the hysteretic deformations vector,  $\mathbf{I}_b = \left(1,1,1,1\right)^T$, $\mathbf{M} = \textrm{diag}\left( m_1, m_2, m_3, m_4\right)$ is the mass matrix,
\begin{equation}{\label{BW_simp_mats}}
    \mathbf{K} = 
    \begin{pmatrix*}
    D_1K_1 & -D_2K_2 & 0 & 0 \\ 0 & D_2K_2 & -D_3K_3 & 0 \\ 0 & 0 & D_3K_3 & -D_4K_4 \\ 0 & 0 & 0 & D_4K_4
    \end{pmatrix*}; \quad
    \mathbf{C} = 
    \begin{pmatrix*}
    c_1+c_2 & -c_2 & 0 & 0 \\ -c_2 & c_2+c_3 & -c_3 & 0 \\ 0 & -c_3 & c_3+c_4 & -c_4 \\ 0 & 0 & -c_4 & c_4
    \end{pmatrix*};
\end{equation}
 and $\mathbf{K}_{\textrm{lin}}$ is the linear stiffness matrix which has the same matrix structure of $\mathbf{C}$ with $K_j \text{ replacing } c_j,\\\forall j=1,\ldots,4$. The hysteretic deformations are governed by:
\begin{equation}{\label{BW_simp_dof_r}}
    \dot{r}_j = \frac{1}{D_j}\left(\dot{y}_{\textrm{drift},j} - \beta_j|\dot{y}_{\textrm{drift},j}|{|r_j|}^{n_j-1}r_j - \gamma_j\dot{y}_{\textrm{drift},j}{|r_j|}^{n_j} \right) \quad ,\text{for}\ j=1,\ldots,4
\end{equation}
where $D_j$ is the yield displacement of the BW element between the $(j-1)$th and $j$th DOFs, with $\beta_j$, $\gamma_j$ and $n_j$ being the corresponding BW parameters, and $y_{\textrm{drift},j}$ is the drift between the $(j-1)$th and $j$th DOFs, defined as: $y_{\textrm{drift},j} = y_{j}$, for $j=1$, and $y_{\textrm{drift},j} = y_{j} - y_{j-1}$, for $j=2,\ldots,4$. The values of the structural parameters are provided in Table~\ref{Table: BW parameters}, and the ratio of post-yield stiffness~to initial stiffness $\alpha = 0.1$ is the same for all BW elements. 

The input base excitation $\Ddot{u}_b\left(t\right)$ is modelled as a uniformly modulated non-stationary process~\cite{priestley1965evolutionary}. Its power spectrum is adopted~\cite{deodatis1996non} as:
\begin{equation}
    S_{\ddot{u}_b}\left(\omega,t\right) = S_{KT}\left(\omega\right)S_C\left(\omega\right)\left(A(t)\right)^2
\end{equation}
where $S_{KT}$ is the Kanai-Tajimi spectrum, $S_C$ is the Clough-Penzien filter~\cite{clough1993dynamics}, and $\left(A(t)\right)^2$ is an exponential modulating function~\cite{der1989evolutionary,hernandez2024lower}, which are given as:
\begin{equation}
    S_{KT}\left(\omega\right) = S_0 \frac{1 + 4\zeta_g^2\left(\frac{\omega}{\omega_g} \right)^2}{\left[1 - \left(\frac{\omega}{\omega_g} \right)^2\right]^2 + 4\zeta_g^2\left(\frac{\omega}{\omega_g} \right)^2}; 
    \quad
    S_C\left(\omega\right) = \frac{\left(\frac{\omega}{\omega_f} \right)^2}{\left[1 - \left(\frac{\omega}{\omega_f} \right)^2\right]^2 + 4\zeta_f^2\left(\frac{\omega}{\omega_f} \right)^2};
    \quad
    (A(t))^2 = te^{-bt}
\end{equation}
The parameters $\omega_g = 10$ rad/s, $\zeta_g = 0.4$, $\omega_f = 1$ rad/s, $\zeta_f = 0.6$ are adopted considering medium stiffness soil~\cite{der1991response}, and the parameters $S_0 = 1/f_s$ and $b = 0.2$, where $f_s$ is the sampling frequency of $\Ddot{u}_b\left(t\right)$. Any realization of $\Ddot{u}_b\left(t\right)$ from its spectrum $S_{\ddot{u}_b}$ can be obtained through the spectral representation theory~\cite{shinozuka1972digital}. 

\begin{figure}[h!]
  \centering
  \includegraphics[width=1\textwidth]{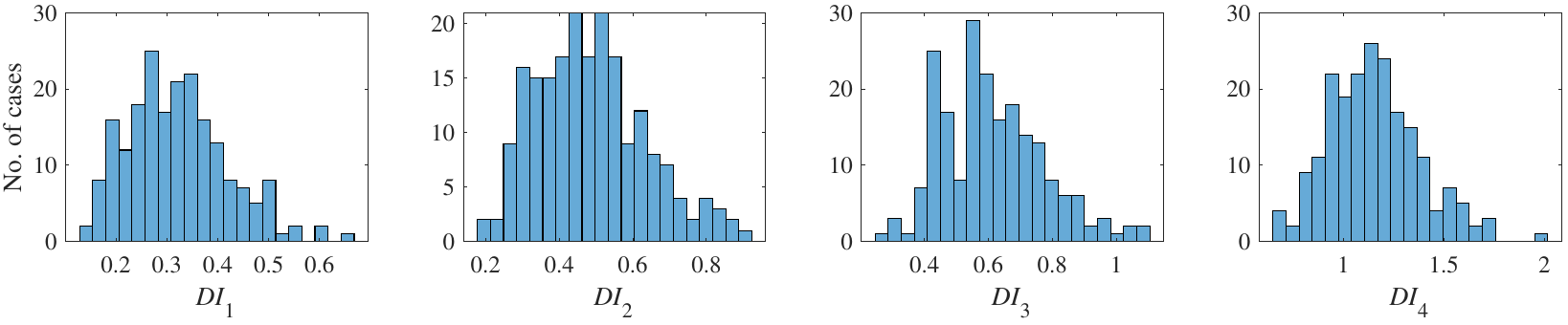}
  \caption{Variation of Park-Ang damage indices $DI_j, \forall j = 1,\ldots,4$ for 200 realizations of $\Ddot{u}_b\left(t\right)$}
  \label{Fig: DI true variation}
\end{figure}

In this example, a Monte Carlo simulation of system identification is carried out with 200 different realizations of $\Ddot{u}_b\left(t\right)$. Each realization is generated with sampling frequency $f_s = 100$ Hz, and is ensured to have a peak ground acceleration less than 0.4 g. For each applied $\Ddot{u}_b\left(t\right)$, the responses at each DOF are computed using the fourth-order Runge-Kutta integration scheme with a time step of 0.001 seconds. Considering the structure to represent a four-story shear building, Park-Ang damage index is computed for each story by using the maximum story drift instead of $y_{\textrm{max}}$ in Eq.~\ref{Park_Ang}, and is denoted as $DI_j$ for the $j$th story. The variations of damage indices for all stories across all 200 simulations are shown in Fig.~\ref{Fig: DI true variation}. It is to be noted that $DI_4$ is mostly greater than 1, which indicates a collapse damage state; this occurs because of very fat hysteresis loops being developed by the BW element of story 4 (between DOFs 3 and 4), which is intentionally designed in this manner to illustrate the patterns of erroneously identified BW parameters in such cases.

\begin{table}[h!]
\centering
\caption{Structural parameters for the four DOFs Bouc-Wen chain model}
\label{Table: BW parameters}
\begin{tabu}{c|ccccccc}
\tabucline[1.5pt]{-}
\textbf{Parameter} & $m_j(\textrm{kg})$ & $K_j(\textrm{N/m})$ & $c_j(\textrm{Ns/m})$ & $D_j(m)$ & $\beta_j$ & $\gamma_j$ & $n_j$ \\
\hline
DOF 1 & 1 & 240 & 1.5 & 0.06 & 2.5 & 1 & 2\\
DOF 2 & 1 & 200 & 1.5 & 0.06 & 3 & 1.5 & 2\\
DOF 3 & 1 & 160 & 1.5 & 0.06 & 4 & 2 & 2\\
DOF 4 & 1 & 90  & 1.5 & 0.06 & 5 & 3 & 2\\
\tabucline[1.5pt]{-}
\end{tabu}
\end{table}

For system identification of the structure, joint state-parameter estimations are carried out in all the 200 simulations, utilizing the Truncated Unscented Kalman filter (TUKF) algorithm \cite{kundu2025truncated}. Absolute accelerations from all four DOFs, corrupted with 10\% root mean squared additive white Gaussian noise, are considered as the measurements from the system. All the ground motion inputs are also corrupted with the same noise level to simulate the measured input. The quantities $\mathbf{M}$, $\alpha$ and $D_j, \forall j = 1,\ldots,4$, are assumed to be known, and the augmented state vector $\mathbf{x}(t) = \left( \mathbf{y}^T, \dot{\mathbf{y}}^T, \mathbf{r}^T, \boldsymbol{\Theta} \right)^T$ is estimated, where $\boldsymbol{\Theta} = \left(K_1, K_2, K_3, K_4, c_1, c_2, c_3, c_4, \beta_1, \beta_2, \beta_3, \beta_4, \gamma_1, \gamma_2, \gamma_3, \gamma_4, n_1, n_2, n_3, n_4 \right)$. As the parameters in $\boldsymbol{\Theta}$ are constant, the evolution of any $\Theta_i\in\boldsymbol{\Theta}$ is considered as $\dot{\Theta}_i = 0$. The constraints utilized in the TUKF are:
\begin{equation}{\label{BW_simp_constr}}
    \begin{split}
    K_j\geq0; \quad
    c_j\geq0; \quad 
    \beta_j+\gamma_j\geq0; \quad
    \beta_j-\gamma_j\geq0 \quad \textrm{and} \quad
    n_j\geq1
    \end{split}
\end{equation}
which ensure physical meaningfulness of the parameters. The process noise matrix, the initial state estimate and the initial covariance matrix are suitably selected, as discussed in Kundu and Mukhopadhyay~\cite{kundu2025truncated}.

\begin{figure}[h!]
  \centering
  \includegraphics[width=0.8\textwidth]{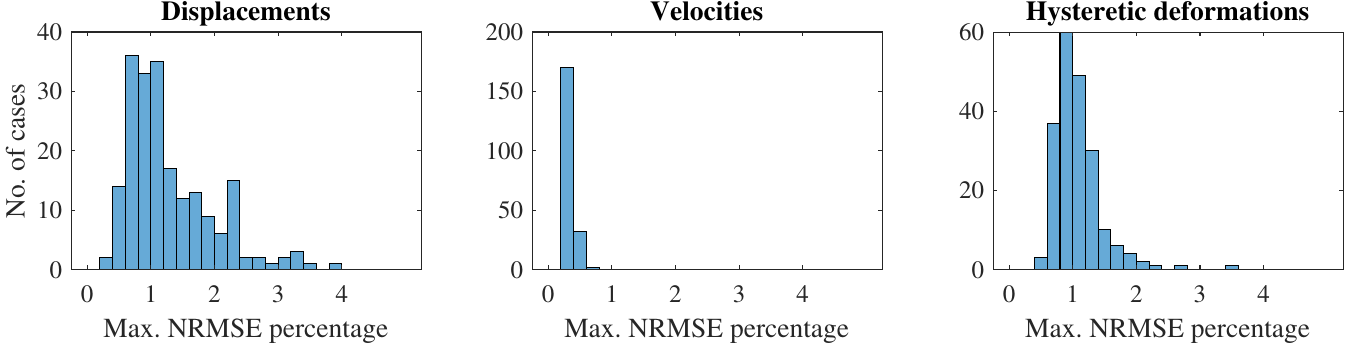}
  \caption{Variation of maximum NRMSE percentages of estimated displacements, velocities, and hysteretic deformations, for 200 realizations of $\Ddot{u}_b\left(t\right)$.}
  \label{Fig: states NRMSE}
\end{figure}

The NRMSE percentages, $\Delta_{\textrm{NRMSE,\%}}(\cdot)$, of the estimates of states $\mathbf{y}, \dot{\mathbf{y}}$ and $\mathbf{r}$ are computed with respect to their corresponding true values, by using Eq.~\ref{NRMSE}. For each simulation, the maximum NRMSE percentages of all floor displacements $y_j,\forall j=1,\ldots,4$, all floor velocities $\dot{y}_j,\forall j=1,\ldots,4$ and all hysteretic story deformations $r_j,\forall j=1,\ldots,4$, are calculated separately, and their variations across all 200 simulations are shown in Fig.~\ref{Fig: states NRMSE}. As the errors are quite low, it can be inferred that all these dynamic states are being estimated with high accuracy. 

\begin{figure}[h!]
  \centering
  \includegraphics[width=0.99\textwidth]{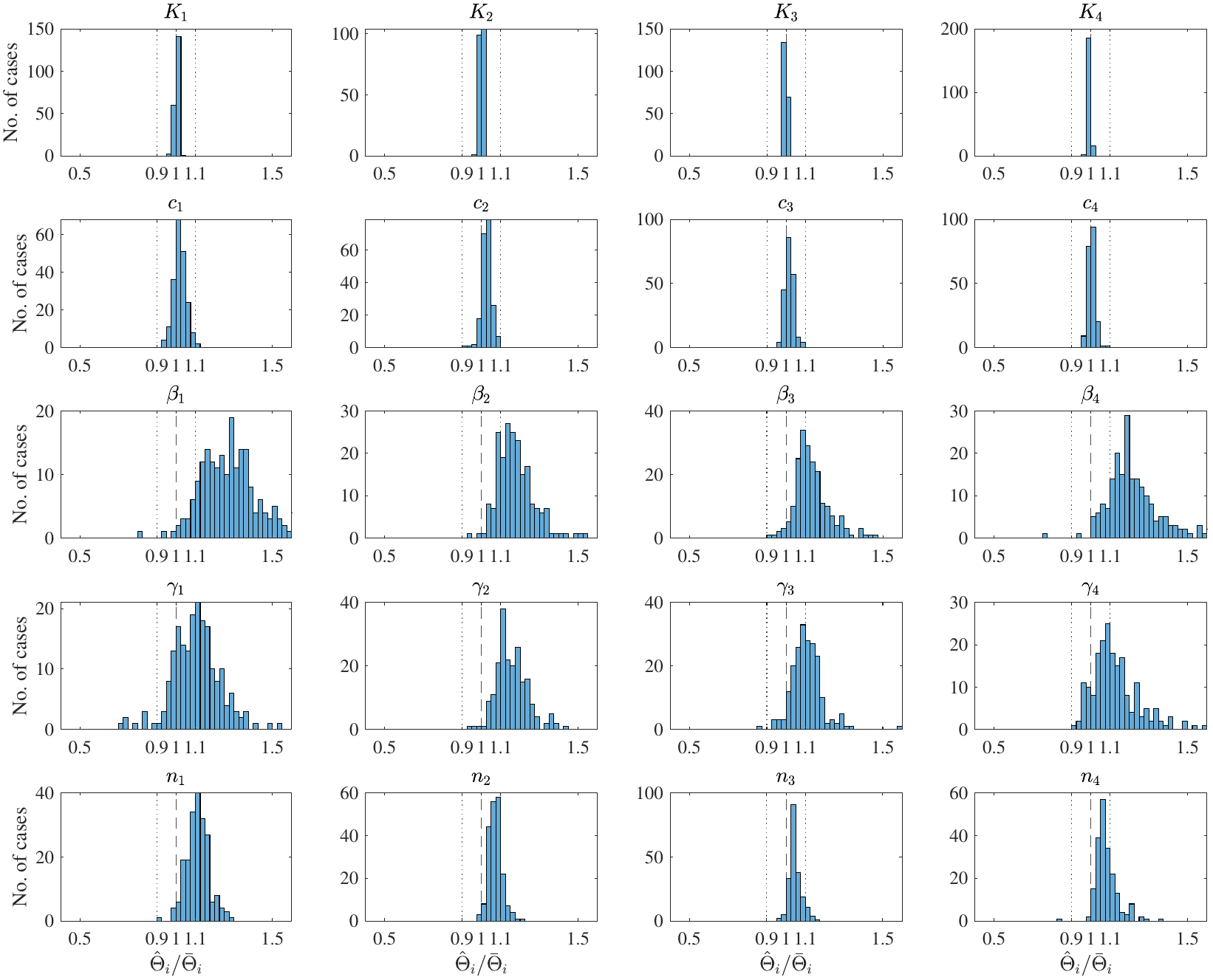}
  \caption{Variation of normalized parameter estimates, for 200 realizations of $\Ddot{u}_b\left(t\right)$.}
  \label{Fig: parameter error}
\end{figure}

For assessing the estimates of any parameter $\Theta_i\in\boldsymbol{\Theta}$, its converged estimate $\hat{\Theta}_i$ is considered and normalized as $\hat{\Theta}_i/\bar{\Theta}_i$, where $\bar{\Theta}_i$ is the true value of $\Theta_i$. The variations of the normalized parameter estimates $\forall \Theta_i\in\boldsymbol{\Theta}$ across all 200 simulations are shown in Fig.~\ref{Fig: parameter error}. It can be seen that the parameters $K_j$ and $c_j, \forall j = 1,\ldots,4$, are estimated accurately in all cases, with all estimates of $K_j$ having less than 3\% error and most of the estimates of $c_j$ having less than 5\% error. However, for the BW parameters, the non-uniqueness in their estimates is clearly evident from the substantial deviations of the estimates from their true values. Particularly for $\beta_j$ and $\gamma_j$, the errors in their estimates can easily cross 10\%, and can even reach as high as more than 50\% for most of these parameters.

\begin{figure}[h!]
  \centering
  \includegraphics[width=0.95\textwidth]{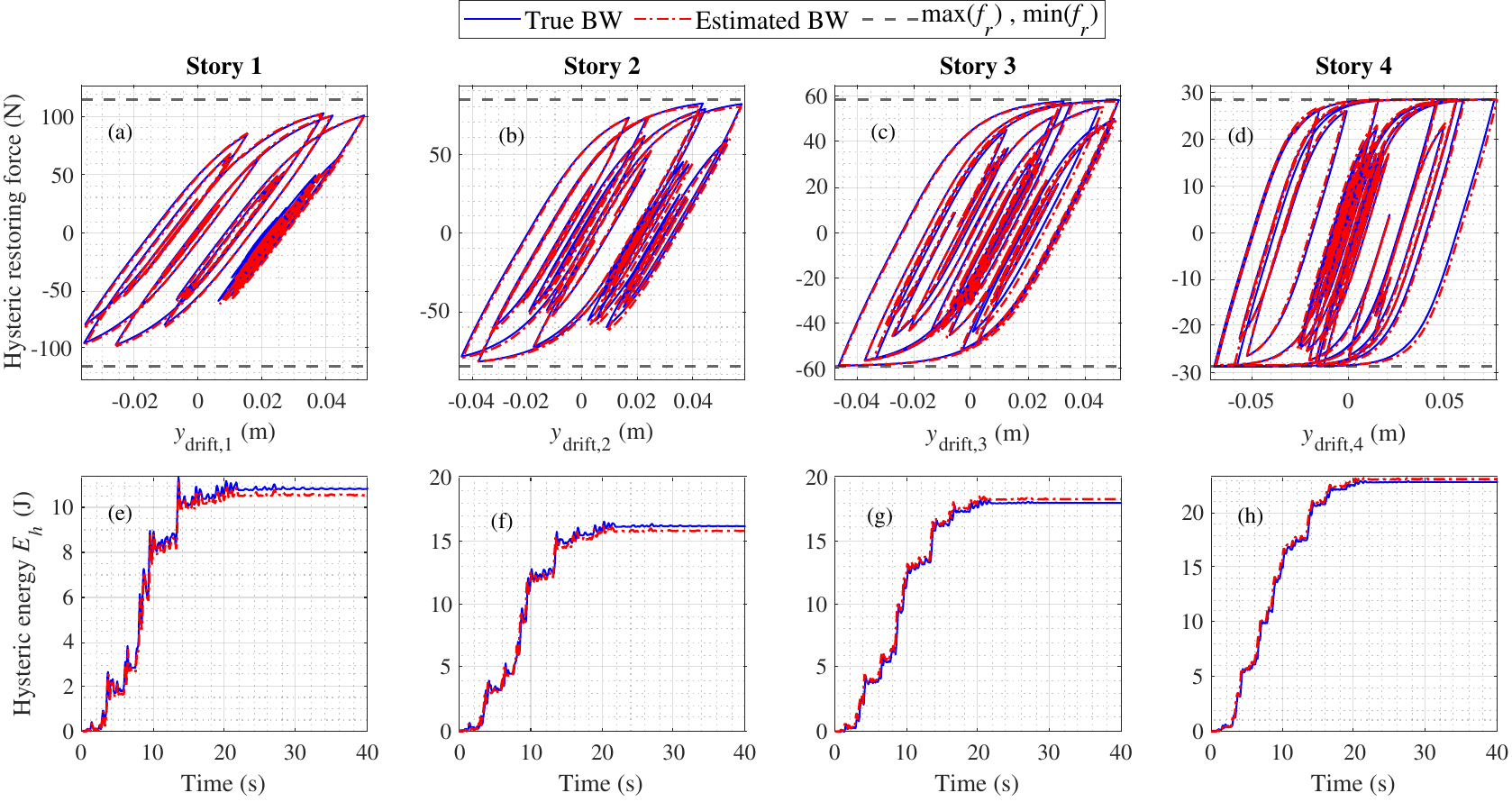}
  \caption{Comparison of hsyteretic behaviours obtained with true and incorrectly estimated BW parameters: (a-d) hysteretic force-deformation behaviours of BW elements of all stories, and (e-g) hysteretic energy dissipated $E_h$ of all stories.}
  \label{Fig: Estimated BW pars}
\end{figure}

For a detailed examination, one of the 200 simulations is chosen. The percentage error of the parameter estimates, denoted by $\Delta_{\%}(\Theta_i)$, are shown in Table~\ref{Table: parameter deviation errors} for this simulation. As expected, the errors in the BW parameter estimates are quite high, while those of $K_j$ and $c_j$ are quite low. Using the incorrectly estimated BW parameters and keeping the parameters $K_j$ and $c_j$ unchanged, the four storied building is again simulated, under the same excitation corresponding to this simulation, and the obtained hysteretic force-deformation behaviour $f_r$ versus $y_{\textrm{drift},j}$, for the BW elements of all stories, are compared with their true hysteretic behaviours in Fig.~\ref{Fig: Estimated BW pars}. The hysteretic energy dissipated $E_h$ obtained with the true and the alternate (incorrectly estimated) BW parameters are also compared in the figure. The alternate set of identified BW parameters can be seen to result in hysteretic behaviours which are very similar to those for the true BW parameters. This also leads to the alternate parameters giving dynamic responses which are very similar to those obtained with the true parameters. It can be verified that these parameters give low values of $\varepsilon_1(r)$ and $\varepsilon_2(r)$ metrics, as expected following the discussions in the previous sections. 

\begin{table}[h!]
\centering
\caption{Percentage errors of parameter estimates}
\label{Table: parameter deviation errors}
\begin{tabu}{cc|cc|cc|cc|cc}
\tabucline[1.5pt]{-}
$\Theta_i$ & $\Delta_{\%}(\Theta_i)$ & $\Theta_i$ & $\Delta_{\%}(\Theta_i)$ & $\Theta_i$ & $\Delta_{\%}(\Theta_i)$ & $\Theta_i$ & $\Delta_{\%}(\Theta_i)$ & $\Theta_i$ & $\Delta_{\%}(\Theta_i)$ \\
\hline
$K_1$ & 1.16 & $c_1$ & -0.28 & $\beta_1$ & 55.71 & $\gamma_1$ & 12.87 & $n_1$ & 20.71 \\
$K_2$ & 0.80 & $c_2$ & 0.88 & $\beta_2$ & 42.81 & $\gamma_2$ & 27.60 & $n_2$ & 17.26 \\
$K_3$ & 0.78 & $c_3$ & 3.27 & $\beta_3$ & 12.47 & $\gamma_3$ & 12.80 & $n_3$ & 5.60 \\
$K_4$ & -0.69 & $c_4$ & 0.42 & $\beta_4$ & 39.88 & $\gamma_4$ & 27.35 & $n_4$ & 13.73 \\
\tabucline[1.5pt]{-}
\end{tabu}
\end{table}

\begin{figure}[h!]
  \centering
  \includegraphics[width=1\textwidth]{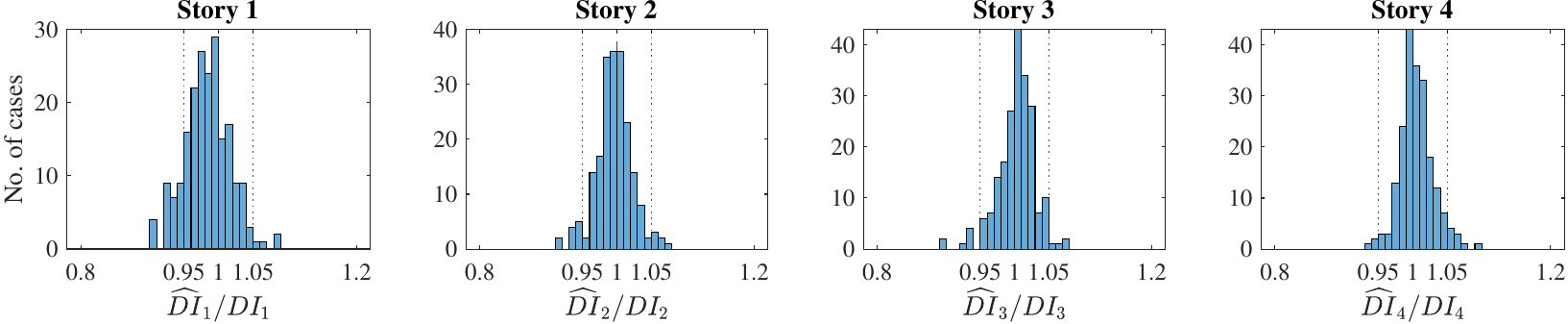}
  \caption{Variation of normalized Park-Ang damage index estimates, across 200 realizations of $\Ddot{u}_b\left(t\right)$.}
  \label{Fig: DI error}
\end{figure}

Regarding damage assessment, as the dynamic responses and the hystertic energies are accurately estimated, the Park-Ang damage indices are also estimated accurately. The true damage indices in this simulation are: $DI_1 = 0.332$, $DI_2 = 0.496$, $DI_3 = 0.614$, $DI_4 = 1.267$, and the estimated damage indices are: $\widehat{DI}_1 = 0.329$, $\widehat{DI}_2 = 0.489$, $\widehat{DI}_3 = 0.620$, $\widehat{DI}_4 = 1.285$. The accuracy of the $DI_j$ estimates is observed in all the 200 simulations, with most estimates having within 5\% error. This is evident from Fig.~\ref{Fig: DI error}, where the variations of the normalized estimates of the damage index $\widehat{DI}_j/DI_j, \forall j = 1,\ldots,4$, across all 200 simulations are shown.

In Fig.~\ref{Fig: Estimated BW pars}, for the BW element of story 4 (between DOFs 3 and 4), the hysteretic force can be seen to reach its maximum and minimum values, at $r_4 = \left(\beta_4 + \gamma_4\right)^{-\frac{1}{n_4}}$ and $r_4 = -\left(\beta_4 + \gamma_4\right)^{-\frac{1}{n_4}}$, respectively, and sometimes sustaining it for considerable amounts of time. This is somewhat similar to the case in Section~\ref{Section: effect of input}, where the SDOF system was subjected to the sinusoidal ground excitation. Therefore, the alternate estimated BW parameters for the BW element between DOFs 3 and 4 give $\epsilon_1 = -0.023$, a very low value. The values of $\epsilon_1$ are relatively higher for the alternate parameters estimated for the other BW elements, where the corresponding hysteretic forces do not reach their extreme levels.

The four DOFs hysteretic system can also be chosen to be modelled without utilizing any $D_j, \forall j = 1,\ldots,4$, in Eqs.~\ref{BW_simp_mats} and \ref{BW_simp_dof_r} i.e., by considering $D_j=1, \forall j = 1,\ldots,4$. In such a case, it can be inferred from Eq.~\ref{BW equivalent} that the resulting equivalent BW parameters $\beta_j$ and $\gamma_j$ of this system will be $(1/0.06)^2 = 277.78$ times their corresponding values in Table~\ref{Table: BW parameters}. With such large magnitudes of $\left(\beta_j + \gamma_j\right),\ \forall j = 1,\ldots,4$, there exist alternate BW parameters having significant deviations from their true values, such as $\Delta_{\%}(\beta_j)>500\%$, which still result in hysteretic behaviours and dynamic responses very similar to those from the true parameters. Significant deviations of the alternate parameters from their true values can also be observed for cases when $\left(\beta_j + \gamma_j\right)\ll 1$. During system identification with the TUKF (or any other method) in these cases, significantly different estimates of the BW parameters may get identified instead of the true parameter values.



\section{Conclusion}

This work highlights the aspect of potential non-uniqueness in the vibration-based system identification of the BW parameters $\beta, \gamma$ and $n$ in hysteretic structures, which is owed to the insensitivity of these parameters towards the dynamic responses and force-deformation behaviour. The insensitivity of these parameters is established by showing the existence of various sets of alternate BW parameters which produce very similar hysteretic force behaviours and dynamic responses. During system identification using measured dynamic responses, this may lead to potential non-uniqueness, i.e., the identification of multiple sets of different BW parameters, all of which can result in very similar hysteretic restoring force behaviours, dynamic responses and damage estimations.

First, a scaled rate of change of the hysteretic force with displacement is derived, which is solely dependent on the hysteretic deformation and BW parameters, making the characterization of the force-deformation behaviour more tractable. Then, the BW parameters are varied to create alternate sets of parameters, which produce new rates of change of hysteretic force. With the objective of finding alternate BW parameters which vary substantially from the true BW parameters, and yet result in the differences between the new and original rates of change of hysteretic force to be very low over the entire valid range of hysteretic deformation, the approximate expressions for these differences are derived. Several metrics are developed for quantifying these approximate differences between the new and original rates of change of hysteretic force, and a wide range of alternate BW parameters are considered. It is shown that multiple combinations of alternate parameters, which vary substantially from the true parameters, result in very low values of the mentioned metrics. It is also seen that, for high magnitudes of $\left(\beta + \gamma\right)$ as well as for magnitudes of $\left(\beta+\gamma\right)\ll1$, much larger deviations from the true values for $\beta$ and $\gamma$ of the alternate parameters can be found, which give low values of the metrics. As the modelling of BW elements can be deliberately done with high or low magnitudes $\left(\beta + \gamma\right)$, these practices should be avoided as significantly different alternate BW parameters producing very similar hysteretic behaviours will exist. 

For numerical validation of alternate BW parameters providing similar hysteretic force-deformation behaviour and dynamic responses, an SDOF lumped mass hysteretic system, subjected to El Centro ground motion, is utilized. The results show that, for the alternate BW parameters, whose rates of change of hysteretic force with deformations are close to that of the original BW parameters, the NRMSE errors of the dynamic responses and restoring forces are typically quite low (such as <1\% error). For such alternate parameters, damages estimated through Park-Ang damage indices are also very close to that for the original BW parameters. A sinusoidal ground excitation is also considered, which causes much fatter hysteretic loops, with the hysteretic force staying at its extreme levels for considerable amounts of time. In this case, the range of alternate BW parameters is comparatively narrower, but it exists nonetheless. The regions of the alternate BW parameters depend on the extent of the hysteresis action being developed by the input excitation. For lower levels of excitations, the regions of alternate BW parameters tend to be relatively larger than the regions corresponding to higher levels of excitations.

Finally, the insensitivity-induced potential non-uniqueness in the system identification of the
BW parameters in structures modelled with BW restoring force mechanisms is illustrated with a Monte Carlo simulation. A four DOF mass-spring-dashpot chain-type model with BW restoring force elements, subjected to 200 base excitations generated from a Kanai-Tajimi type spectrum, is considered. Assuming the availability of noisy absolute acceleration measurements from all DOFs of the system, joint state-parameter estimations are performed with the TUKF algorithm. It is seen that all the dynamic states and all parameters, apart from the BW ones, are being estimated accurately. It is observed that the incorrectly estimated BW parameters belong to the set of alternate BW parameters producing scaled rates of change of hysteretic force with displacement very close to those of true BW parameters. On utilizing the incorrectly estimated alternate parameters for modelling the four DOFs systems, it is seen that the restoring force behaviours, the dynamic responses, and the damage indices of the incorrectly estimated model (with incorrect BW parameters) are very similar to those of the correct model (with true BW parameters).

From this study, it can also be concluded that, during system identification of structures modelled with BW restoring force mechanisms, instead of trying to accurately obtain the BW parameters (which may not be possible due to the potential non-uniqueness highlighted in this paper), the focus should rather be on the accurate estimation of the hysteretic restoring force, along with other sensitive parameters, such as damping and stiffness, the dynamic states, and the damage index, for performing reliable damage assessment.

\section{Acknowledgments}

Financial support for this work has been provided by the Science and Engineering Research Board (SERB, DST, India), under grant number SERB/CE/2023789. The financial support is gratefully acknowledged.

\printcredits

\bibliographystyle{model1-num-names}

\bibliography{cas-refs}


\end{document}